\documentclass{article}
\usepackage{graphicx}
\def\drcaption#1{
    \footnotesize Figure~\thefigure\quad #1}%
\newenvironment{fig}[6]{%
\begin{minipage}[t]{#1 pt}
\begin{picture}(#1,#2)
\put(0,0){\epsfig{file=#3,width=#1 pt, height=#2 pt}}
#4
\end{picture}
\refstepcounter{figure}
\drcaption{#5}%
\label{#6}
\vspace{6pt}}
{\end{minipage}}
\newenvironment{drfigs}{\par \ignorespaces}
{\par\vspace{12pt}\ignorespaces}%
\def\drsetfig#1#2#3#4{
\noindent%
\begin{minipage}[b]{#1}    
\begin{center}    
\leavevmode\epsfig{file=#2,width=#1} 
\end{center}   
\refstepcounter{figure}    
\drcaption{#3}\label{#4}  
\end{minipage}\ignorespaces}%
\def\dr2setfig#1#2#3#4#5{
    \noindent%
    \begin{minipage}[b]{370 pt}
    \begin{center}
    \leavevmode
\epsfig{file=#2,width=#1} \hspace{20pt} \epsfig{file=#3,width=#1} 
    \refstepcounter{figure}
    \drcaption{#4}\label{#5}
    \end{center}
    \end{minipage}\ignorespaces}
\newif\ifmarglabs
\marglabsfalse
\def\drlabel#1{\ifmarglabs
\marginpar{\scriptsize label: \\ #1} \else \relax\fi}
\usepackage{epsfig}
\newcommand{\sA}{{\cal A}}

\newcommand{\sC}{{\cal C}}

\newcommand{\sF}{{\cal F}}

\newcommand{\sJ}{{\cal J}}
\newcommand{\sJt}{\widetilde{\sJ}}

\newcommand{\Fo}{F_{0}}
\newcommand{\Fs}{F_{s}}
\newcommand{\Fmu}{F_{\mu}}
\newcommand{\Fcrit}{F_{{\rm crit}}}
\newcommand{\Fm}{F_{\mu}}

\newcommand{\Ie}{I_{e}}
\newcommand{\Ize}{I_{0}}
\newcommand{\Ion}{I_{1}}
\newcommand{\Itw}{I_{2}}

\newcommand{\In}{I_{n}}
\newcommand{\Ism}{I_{m}}

\newcommand{\Jm}{J_{m}}

\newcommand{\Kmean}{K_{m}}
\newcommand{\nze}{n_{0}}

\newcommand{\thetaon}{\theta_{1}}
\newcommand{\thetatw}{\theta_{2}}
\newcommand{\thetae}{\theta_{e}}
\newcommand{\phin}{\phi_{n}}
\newcommand{\phie}{\phi_{e}}
\newcommand{\phim}{\phi_{m}}
\newcommand{\psie}{\psi_{e}}

\newcommand{\omegae}{\omega_{e}}

\newcommand{\Kbb}{\overline{K}_{m}}
\newcommand{\Kbr}{\overline{K}_{R}}
\newcommand{\Kr}{K_{R}}

\newcommand{\omo}{\Omega_{0}}

\newcommand{\gb}{\overline{g}}
\newcommand{\gt}{\tilde{g}}

\newcommand{\bI}{{\bf I}}
\newcommand{\bn}{{\bf n}}

\newcommand{\bp}{{\bf p}}

\newcommand{\br}{{\bf r}}

\newcommand{\bz}{{\bf z}}
\newcommand{\bZ}{{\bf Z}}

\newcommand{\vsp}[1]{\rule{0pt}{#1 pt}}
\newcommand{\hsp}[1]{\hspace*{#1 pt}}

\newcommand{\disp}{\displaystyle}

\newcommand{\Pion}{P_{i}}
\newcommand{\Pfit}{P_{i}^{\rm fit}}

\newcommand{\cd}{\partial}

\newcommand{\pad}[2]{\frac{\cd #1}{\cd #2}}

\newcommand{\ket}[1]{\left| #1 \right\rangle}
\newcommand{\wave}[2]{ \left\langle #1 \left| \right. #2 \right\rangle}
\newcommand{\mean}[1]{ \left\langle #1 \right\rangle}
\newcommand{\range}[4]{#1=#2,\,#3,\cdots,#4}

\begin{document}


\title{The effect of parallel static and microwave electric fields on
excited hydrogen atoms}
\author{D Richards\\ Faculty of Mathematics\\
Open University\\
Walton Hall\\
Milton Keynes, MK7 6AA.}
\date{\today}
\maketitle

\begin{abstract}
Motivated by recent experiments we 
analyse the classical dynamics of a hydrogen atom in parallel
static and microwave electric fields.
Using an appropriate representation and averaging approximations we show
that resonant ionisation is controlled by a separatrix,
and provide necessary conditions for a dynamical resonance to affect 
the ionisation probability.

The position of the dynamical resonance is computed
using a high-order perturbation series, and estimate its radius 
of convergence. We show that
the position of the dynamical resonance does not
coincide precisely with the ionisation maxima, and that the field switch-on 
time can dramatically affect 
the ionisation signal which, for long switch times, reflects the shape
of an incipient homoclinic. Similarly, the resonance ionisation 
time can reflect the time-scale
of the separatrix motion, which is therefore longer than 
conventional static field Stark ionisation. We explain why these effects 
should be observed in the quantum dynamics.\\[3pt]
PACs: 32.80.Rm, 33.40.+f, 34.10.+x, 05.45.Ac, 05.45.Mt 
\end{abstract}

\newpage
\section{Introduction}
A strong electromagnetic fields can perturb an atom in many unexpected and 
complicated ways that are difficult to understand. If the atom is initially
in an excited state  usually
a large number of unperturbed bound states are coupled,
making the numerical solution of Schr\"{o}dinger's equation difficult.
Moreover, the corresponding classical dynamics is normally partially
chaotic --- meaning that there are
both unstable and stable orbits close to the initial unperturbed torus --- and
the wave function mimics this behaviour, see for instance Leopold
and Richards 1994 and Richards 1996b, thus making the interpretation of
numerical solutions difficult.

The investigation of the effects of strong periodic fields on an excited
atom dates from the original experiments of Bayfield and Koch (1974) which
showed that a relatively weak field could produce a multiphoton transition
into the continuum, contrary to the received wisdom of quantal perturbation
theory. The subsequent  history of this interesting problem is told by Koch
(1990). In 1974 conventional quantal theory required high-order perturbation
theory to describe the 80 photon jumps to the continuum of this early 
experiment. This was, and remains, an impossible calculation, but theory
was rescued by Delone {\em et al} (1978) who proposed that the classical
ionisation mechanism involved diffusion of the electron through atomic 
states highly perturbed by the field. A year later Meerson {\em et al} (1979) 
proposed a different classical diffusion approximation. In the same year
Leopold and Percival (1979) 
used a classical Monte-Carlo trajectory method
to estimate classical ionisation probabilities and
obtained qualitative agreement with experiment. In 1985 the experiment was
repeated with better control of all important parameters and the comparison
between these results and a classical Monte-Carlo simulation, 
van Leeuwen {\em et al} (1985), showed remarkable agreement.

Since then our understanding of the dynamics of this type of system 
and the relationship between classical and quantal solutions in 
different parameter regimes has developed. For instance we now know
that the scaled frequency --- the ratio between the 
driving  frequency and the Kepler frequency of the initial unperturbed 
motion --- is one of the most significant parameters and that there are six 
separate scaled frequency regions in which the dynamics has quite different
characteristics, see for instance Koch and van Leeuwen (1995).

Linearly polarised fields with low scaled-frequencies 
were considered in Richards {\em et al} (1989) and there 
it was found that at particular scaled frequencies quantal
effects were important due to resonances between two adiabatic states,
see also Dando and Richards (1993). Low frequency elliptically
polarised fields, Bellermann {\em et al} (1996, 1997) and Koch and 
Bellermann (2000), showed the existence of complicated resonance structures 
that, on the other hand, can be explained using classical dynamics, 
Richards (1997).
When the scaled frequency is close to unity the main classical resonance island
plays a dominant and similar role in both the classical and quantal dynamics,
except at certain frequencies where scars produce differences, 
Leopold and Richards (1994) and Richards (1996b).
At higher scaled frequencies classical dynamics fails, as 
predicted by Casati {\em et al} (1984), and demonstrated by Galvez 
{\em et al} (1988) for linearly polarised fields and, for elliptically 
polarised fields, by Wilson (2003).
These different behaviours  have been the subject of several reviews 
Jensen {\em et al} (1991), Koch (1990, 1995), Koch and van Leeuwen (1995).

In this paper we consider the effect of strong, parallel static and 
oscillatory fields
on an excited hydrogen atom, the work being motivated by recent experiments
of Professor Koch's group. These two fields affect the system in complicated
ways. Roughly, the static field splits the hydrogen degeneracy introducing
another frequency to the system with which the external field can resonate.
Classically this means that the Kepler ellipse moves periodically
and this motion can resonate with the driving field. Such 
resonances can enhance the ionisation probability so a significant part of
this paper is devoted to understanding the dynamics of these resonances and
the mechanism causing enhanced ionisation. Preliminary experimental
results and some comparisons with classical Monte-Carlo calculations are
given in Galvez {\em et al} (2000); a more detailed discussion of the 
experimental method with more extensive results and comparisons with classical
calculations is provided in
Galvez  {\em et al} (2004), subsequently referred to as I.

In these experiments excited
hydrogen atoms are subjected to strong parallel static and periodic fields, so
the external force on the electron is $(\Fs+\Fmu\cos\Omega t) \hat{\bz}$:
the accurate theoretical description of this system present a challenge. 
The main observed effect of this field combination is to produce 
an ionisation probability that, for fixed $\Fmu$ and $\Omega$, rises 
steadily as $\Fs$ is increased, but which is punctuated by a series of 
approximately 
equally spaced sharp local maxima, see for instance Galvez {\em et al} 
(2000, figure~1) and figure~\ref{f:1} of this paper. Each local maximum
is produced by a resonance between the periodic part of the perturbation,
$\hat{\bz}\Fm\cos\Omega t$, and the mean rotational motion of the 
Kepler ellipse
induced by the static part, $\hat{\bz}\Fs$. These peaks are
relatively narrow but their shapes, widths and heights vary significantly with 
the system parameters. Moreover, they disappear at certain values of
$\Fmu$ and $\Fs$, for reason that are sketched in Galvez {\em et al} 
(2000); similar explanations  are provided by Oks and Uzer 
(2000) and Ostrovsky and Horsdal-Pedersen (2003).

Since publication of the original Galvez {\em et al} (2000)  paper 
we have made detailed comparisons
between the ionisation probabilities of the experiment and classical 
Monte-Carlo simulations, which will be published in Galvez {\em et al} (2004).
These comparisons show remarkable agreement in many instances, but also some
differences, some of which may be attributed to unquantifiable differences
between the assumed substate distribution. In experiment and 
calculations we observe systematic differences between both the positions
of the resonances and of their disappearances and  the
predictions of the simple theories. In an attempt to understand 
these differences more detailed
numerical investigations of the classical dynamics were undertaken and these
calculations show that the dynamics underlying the apparently simple,
averaged results seen in figure~\ref{f:1} is, in fact, very complicated: even
an accurate theoretical prediction of positions of the local maxima in the 
ionisation probability is fraught with difficulties in classical and 
quantum dynamics. Further, it transpires that the nature of the
resonance that causes these maxima is unusual in that the classical resonant 
island moves as the field is switched on: in some circumstances this means that
the classical ionisation probabilities reflect the development of a homoclinic 
tangle, see figure~\ref{f:18a} and~\ref{f:18b}. The size of the quantum 
numbers needed for the quantum dynamics to mimic this behaviour is not known.

This paper has two main aims. First, to provide a theoretical understanding
of the classical dynamics and to isolate those features that
determine the nature of these resonances, for example their positions 
and temporal development. Second we need to derive an approximate
Hamiltonian that leads to a numerically tractable Schr\"{o}dinger equation.
Both aims are achieved by using a representation in which coupling between
basis states is relatively small.

In section~\ref{sec:not} notation is defined and we present some numerical 
results illustrating the theoretical problems that need to be solved.
In section~\ref{sec:thry} the approximations are developed; this 
is algebraically
complicated but, in essence, it is simply a standard use of perturbation
theory and averaging approximations. However,  for two reasons care is 
need with this theory. First, the convergence of the perturbation expansion
needs to be understood because the fields used are strong and, it transpires,
sometimes beyond the radius of convergence of the series. Second, it is shown
that particular observable effects are produced by subsets of terms and it
is necessary to determine their origin because only some subsets can
be computed to high order --- this means that some effects, such as the
resonance positions, can be computed to high order, but others cannot. This
analysis is necessary partly to extend the earlier approaches beyond
first order and partly to show how ionisation occurs, a feature missing
from all earlier theories of this system. Furthermore, from this 
analysis emerges an approximate Hamiltonian that should facilitate a 
quantal calculation.

In section~\ref{sec:res-ion-I} the mechanism connecting the dynamical
resonances, described in section~\ref{sec:thry}, with ionisation
is described. A significant result of this analysis is that the positions
of the dynamical resonance is not precisely the same as the positions
of the local maxima seen in ionisation curves; although small, an estimate
of the difference seems beyond current theory. In this section we determine
some necessary conditions for a dynamical resonance to affect ionisation
and also obtain accurate estimates for the position at which the resonances 
observed by Galvez {\em et al} (2000) disappear, see also Schultz (2003)
and Schultz {\em et al} (2004). We also discuss resonance 
widths and show that for the current experiments, which involve an average
over substates, the width is {\em not} due to the variations of the
resonance position with substate (which is relatively small), but is 
due mainly to non-adiabatic dynamical effects, that are difficult to quantify.

In section~\ref{sec:time-scale} we analyse the time required for a 
resonance to develop and show this to be relatively long. Further, we
observed that the nature of the resonance island means that the classical
ionisation probability can be significantly affected by the field envelope.
In particular in some circumstances the ionisation probability can 
reflect an incipient homoclinic tangle that develops as the initial state
moves slowly through a separatrix.

\section{Notation}
\drlabel{sec:not}\label{sec:not}
The Hamiltonian for a hydrogen atom in parallel 
static and microwave electric fields, as in the experiments of 
Galvez {\em et al} (2000, 2004) (the latter henceforth being referred 
to as I), is derived in Leopold and Richards (1991) and, provided the field 
envelope $\lambda(t)$ changes sufficiently slowly, is given by
\drlabel{eq:2-01}
\begin{equation}
H=\frac{1}{2\mu} \bp^{2} - \frac{e^{2}}{r} + F(t)z, \quad
F(t)=\left( \Fs + \Fmu\cos\Omega t\right)\lambda(t),
\label{eq:2-01}
\end{equation}
where $\mu$ is the atomic reduced mass, $e$ the electron charge and 
$\lambda(t)$ is the envelope function describing the
passage of the atom through the cavity. This Hamiltonian has
azimuthal symmetry so the $z$-component of angular momentum, $\Ism$, is
conserved. For the particular experiments described in~I,
$\lambda(t)$ has the 16-113-16 configuration,
meaning that it rises monotonically from zero to 
unity in 16 field periods, remains constant for 113 periods 
and then decreases monotonically to zero in 16 periods. In all calculations
reported here the initial rise over $N_{a}$ field periods is taken to be
\[
\lambda(t)=x^{2}(2-x^{2}), \quad t=\frac{2\pi N_{a}}{\Omega}x, 
\quad 0 \leq x \leq 1,
\]
and the decrease as the field is switched off has the same shape.
Classical ionisation probabilities are normally insensitive to small changes 
in $\lambda(t)$; exceptions to this rule are discussed in 
section~\ref{sec:envel}.

For excited atoms it is convenient to use units defined by the initial
unperturbed Kepler ellipse which has semi-major axis
\(
a=\Ize^{2}/(\mu e^{2}) 
\) 
and frequency $\omega_{K}=\mu e^{4}/\Ize^{3}$, $\Ize=\nze\hbar$,
where $\Ize$ and $\nze$ are the initial values of principal action and 
quantum number respectively: scaled
units are convenient because the magnitude of most scaled parameters that
produce similar physical effects change little with $\nze$. The scaled 
frequency and field amplitude are defined by
\drlabel{eq:2-02}
\begin{equation}\begin{array}{l}
\disp \omo=\frac{\Omega}{\omega_{K}}=\frac{\Omega\Ize^{3}}{\mu e^{4}}=
\frac{\Omega}{{\rm GHz }}\left( 0.00533757\nze\right)^{3}, \vspace{4pt}\\
\disp \Fo=\frac{a^{2}F}{e^{2}}=\frac{F\Ize^{4}}{\mu^{2}e^{6}}=
\frac{F}{{\rm V/cm}}\left( 0.00373535\nze\right)^{4}.
\end{array}
\label{eq:2-02}
\end{equation}
The scaled time is $t_{0}=\omega_{K}t$ and a scaled action $I$ is $I/\Ize$. 
In the current experiment \mbox{$\Omega=8.105\,$GHz}, 
so $\omo=(0.010722\nze)^{3}$.
In the following we use the symbols
$\Fs$ and $\Fmu$ for both the scaled and actual field amplitudes and $t$ for
scaled and actual time: this misuse of notation avoids a clutter of subscripts
but should not cause confusion because scaled quantities are dimensionless.

\subsection{Some numerical results}
Before describing the theory we show the results of a few classical 
calculations in order to provide the reader with an idea of the features 
that a theory needs to describe. For the present calculations a Monte-Carlo 
method, as described in Abrines and Percival (1966), 
is used in which $N$ initial conditions are chosen from a microcanonical
ensemble: if $M$ of these orbits ionise the estimate of the ionisation 
probability is $\Pion=M/N$. Without stratification the standard deviation of
this estimate is (Hammersley and Handscomb, 1964)
\[
\sigma=\sqrt{\frac{\Pion(1-\Pion)}{N}}, \quad \Pion=\frac{M}{N},
\]
meaning that there is a 68\% and 95\% probability of the true result being 
in the ranges $(\Pion-\sigma,\Pion+\sigma)$ and 
$(\Pion-2\sigma,\Pion+2\sigma)$ respectively.
In the present calculations the sample of initial conditions is stratified 
by dividing the range of each variable into equally probable 
intervals and choosing, from a microcanonical distribution, one point at random
in each sub-interval, see Abrines and Percival (1966). Stratification  
reduces the statistical errors and sample calculations suggest that with
this form of stratification the true value of $\sigma$ is about half the
above estimate. For the numerical integration of Hamilton's equations the
problem associated with the Coulomb singularity is circumvented 
by using the regularisation method described in (Rath and Richards 1988):
numerical integration was performed with the NAG routine D02CAF.

In the first figure is shown a `typical' classical ionisation curve in which
$\Fm$ and $\omo$ are fixed and 
the variation of the ionisation probability, $\Pion$, with the static 
field $\Fs$ is shown: here a microcanonical distribution of substates is 
used. For this illustration we choose $\omo=0.0980$, $(\nze=43)$ and 
$\Fmu=0.10$: the field envelope was 16-80-16 and 1296 orbits, for
each value of $\Fs$,  were used.

\vspace{4pt}

\begin{center}
\drlabel{f:1}
\begin{drfigs}
\drsetfig{288pt}{n43-f1.eps}{Ionisation curve for $\omo=0.0980$, 
$(\nze=43)$, $\Fm=0.1$.}{f:1}
\end{drfigs}
\end{center}

\noindent
The broad features are clear: $\Pion=0$ for $\Fs<0.013$ and $\Pion=1$ for
large $\Fs$, typically \mbox{$\Fs>0.2$}. 
As $\Fs$ increases between these values the steady increase in 
$\Pion(\Fs)$ is punctuated only by sharp local maxima at almost equal
intervals in $\Fs$, at $\Fs=0.0316,\,0.0620$, $0.0916$ and $0.1203$; a close 
inspection of the data shows another small maximum at $\Fs\simeq 0.148$,
marked by the arrow.
The small amplitude undulations seen in this ionisation
curve are assumed to be caused by the statistical errors mentioned above, 
and provide a visual estimate of the magnitude of these errors. The 
variation in the scale-length 
of these oscillations in $\Fs$ is due to variations in $\Fs$-interval used
in the calculations which was
smallest near the local maxima of $\Pion$.

We denote the positions of these maxima in $\Pion(\Fs)$ by $\sF_{s}^{(j)}$. 
Theory associates these maxima with a resonance between the driving field
and the precession of the atomic Kepler-ellipse and we denote the position
of these resonances by $\Fs^{(j)}$. The two field values $\Fs^{(j)}$ 
and $\sF_{s}^{(j)}$ are approximately the same but are not identical, 
which is why it is helpful to use different symbols.

Elementary consideration, 
section~\ref{sec:resham}, show that $\Fs^{(j)}\simeq \omo j/3$ 
(in scaled units), giving 0.0327, 0.0653, 0.0980 and 0.131, so the relative
differences between $\omo j/3$ and $\sF_{s}^{(j)}$ are
3.5\%, 5.3\%, 7\% and 9\% respectively. The more accurate
theory developed in  section~\ref{sec:res-ion-II}, gives 
$\Fs^{(j)}=0.0316$, $0.0623$, $0.0904$ and $0.112$, for $j=1$-4,
with relative differences of 0.5\%, 1.3\% and 7\%, respectively, for 
$j=2$, 3 and 4, with the values for $j=1$ agreeing to three figure accuracy.

The averaged ionisation probabilities disguises a richer and more complicated
behaviour. In the next three figures are shown ionisation probabilities
from a given substate, $\Itw=\Ism=0.2$  --- these scaled actions are defined 
below, equation~\ref{eq:2-08a} and~\ref{eq:2-08b}. Here $\omo=0.0528$, 
($\nze=35$), and 1600 orbits, for each $\Fs$, were used. The arrows point to 
$\Fs=\Fcrit^{(+)}-\Fmu$ where $\Fcrit^{(+)}$ is defined in 
equation~\ref{eq:st30} below.
%

\begin{center}
\drlabel{f:2}
\begin{drfigs}
\drsetfig{360pt}{n35-subs-exact.eps}{
Ionisation probabilities for $\omo=0.0528$, $\Itw(0)=\Ism=0.2$, 
($\Ie(0)=-0.4$), with the 16-50-16 envelope and for various values 
of $\Fm$.}{f:2}
\end{drfigs}
\end{center}

\noindent
At the lowest microwave field, $\Fmu=0.13$, a number of local maxima are 
seen: those labelled $j=1$-$4$ can be associated with a dynamical resonance,
and most are visible for $\Fm=0.14$ and $0.15$. For $\Fm=0.13$ 
there are also a number of other local maxima, at $\Fs=0.0430,\,0.0535$
and $0.0705$ the origin of which is not known, but in 
section~\ref{sec:time-scale} results are presented which suggest that
the maximum at $\Fs=0.0430$ is a non-integer resonance, equivalent to 
$j=2\frac23$; 
In this example $\Pion=0$ for $\Fs<0.016$ and $\Pion=1$ for $\Fs>0.076$, with 
an underlying steady increase in $\Pion(\Fs)$ for $\Fs>0.045$.

For $\Fmu=0.14$ the four labelled maxima of the previous figure
persist but have shifted slightly and the $j=3$ maxima has split: we have 
confirmed that the latter effect is not due to statistical sampling 
errors. In 
other calculations with the shorter rise time of four field periods there is 
no split, suggesting that it is caused by the field envelope: this and other
effects of the field envelope are discussed in the section~\ref{sec:envel}.
The $j=1$ maximum near $\Fs=0.0166$ is lower and the theory developed below
shows that it disappears completely for $\Fmu \simeq  0.147$, 
$\Fs \simeq 0.0161$. 
A new feature in this graph is the large value 
of $\Pion$ at $\Fs=0$, with $\Pion$ falling rapidly to zero as 
$\Fs$ increases to $0.007$. For $\Fs>0.035$ the underlying trend in 
$\Pion$ is a steady increase to unity, but all other structure is 
not understood.

For $\Fmu=0.150$ the three labelled maxima persist, but have shifted and
broadened. Now $\Pion$ is large for $\Fs<0.013$ and $\Fs>0.04$ and there 
is a new maximum at \mbox{$\Fs=0.0192$}, the existence and magnitude of which
depends upon the switch-on time, see figure~\ref{f:5}: we have confirmed
that, at these low frequencies, $\Pion$ is affected insignificantly by the 
switch-off time, provided the total interaction time is sufficiently long.

The numerical details of the resonance positions seen in these figures
are given in table~\ref{t:01}. Here the values of $\sF_{s}^{(j)}$ are computed 
using a grid $\Delta\Fs=0.0001$ and defining $\sF_{s}^{(j)}$ to be the value
of $\Fs$ at which $\Pion(\Fs)$ has a local maximum. If $\Pion > 0.99$ for
a range of fields, the range is quoted and sometimes there is more than
one clear maximum. For fixed $\Fm$ and 
increasing $j$ there are clear systematic differences between these two 
values. It is not known what causes these differences; the breakdown of the 
perturbation expansion used to compute $\Fs^{(j)}$ maybe significant, but also
the discussions in sections~\ref{sec:res-ion-II} 
and~\ref{sec:envel}, show that the relation between the values of 
$\Fs^{(j)}$ and $\sF_{s}^{(j)}$ is far from simple.

\drlabel{t:01}
\begin{table}[htbp]
\caption[]{\small \label{t:01} Values of $\sF_{s}^{(j)}$, computed
by the method described in the text, and $\Fs^{(j)}$ given by
the theory described in section~\ref{sec:res-ion-II}.}

\vspace{4pt}

\begin{center}
\begin{tabular}{| c| cc | cc | cc |} \hline
 & \multicolumn{2}{c}{$\Fm=0.13$} & \multicolumn{2}{c}{$\Fm=0.14$} & 
\multicolumn{2}{c}{$\Fm=0.15$} \\ \hline
$j$ & Monte Carlo & theory & Monte Carlo & theory & Monte Carlo & theory 
\\ 
 & $\sF_{s}^{(j)}$ & $\Fs^{(j)}$ & $\sF_{s}^{(j)}$ & $\Fs^{(j)}$ & 
$\sF_{s}^{(j)}$ & $\Fs^{(j)}$  \\ \hline
 1 & 0.0166 & 0.0166 & 0.0162 & 0.0164 & 0.0158 & 0.0160 \\
 2 & 0.0327 & 0.0330 & 0.0320-0.0323 & 0.0323 & 0.0314-0.0316 & 0.0314 \\
 3 & 0.0486 & 0.0487 & 0.0464,\hspace{2pt}0.0483 & 0.0474 & 0.0455-0.0483 & 
0.0455 \\
 4 & 0.0645 & 0.0631 & 0.0632-0.0641 & 0.0607 &  &  \\ 
\hline \hline
\end{tabular}
\end{center}
\end{table}

\noindent
The regular but not equally spaced series of four minima beyond the arrows, 
in figure~\ref{f:2}, have, at present, no dynamical explanation. Numerical 
evidence, however
suggests that some of this structure is caused by the field switch,
with shorter switch times removing most of the structure and longer times
producing more, see figure~\ref{f:5}: the duration of the 
centre part of the envelope is irrelevant provided it is long enough,
see section~\ref{sec:time-scale}.

In the following two figures we show ionisation curves for the same parameters
used in the right hand panel of figure~\ref{f:2}, but with different envelopes.

\begin{center}
\drlabel{f:5}
\begin{drfigs}
\drsetfig{360pt}{n35-sub-02.eps}{
Ionisation probabilities for $\omo=0.0528$, $\Itw(0)=\Ism=0.2$, 
($\Ie(0)=-0.4$), with $\Fm=0.15$ and various envelopes.}{f:5}
\end{drfigs}
\end{center}

\noindent
For the shorter switch, 4-50-4, $\Pion \sim 1$ for $\Fs<0.014$ and 
$\Fs>0.052$ and there are local maxima only at the $j=1$ and 2 
resonances where $\sF_{s}^{(1)}=0.0158$ and 
$\sF_{s}^{(2)}=0.0315$. For the longer switch, 40-50-40, the above two maxima 
persist, at the same field values, but now
the ionisation curve shows a great deal of other structure, some of which
is sensitive to the switch time. We discuss other effects of the field
switch in section~\ref{sec:envel}.

\section{Theory}
\drlabel{sec:thry}\label{sec:thry}
In the situations of interest here
the field frequency $\Omega$ is small by comparison with the Kepler frequency,
so the scaled frequency $\omo$, equation~\ref{eq:2-02}, is small; typically 
it varies between 0.04 and 0.11 for $\nze$ between 32 and 45 for the 
8.105 GHz cavity. Hence the variation of the field, $F(t)$, is slow 
by comparison with the electron orbital motion and 
an averaging approximation may be used to remove one degree of freedom,
but first it is necessary to choose the correct representation. 
 
The fields encountered here are sufficiently large to 
couple together many states of the field free atom and to
ionise some of these states, so useful theoretical 
descriptions of the experimentally observed signal must include a mechanism 
for ionisation and, ideally, should use a representation in which coupling
between bound states is relatively small.

In order to understand the magnitude of this problem, and to motivate 
the following analysis, we consider the static field ionisation of the 
one-dimensional atom with the Hamiltonian
\[
H_{1}=\frac{1}{2\mu} p^{2}-\frac{e^{2}}{z}+\lambda(t)Fz, \quad z > 0,
\]
where $F$ is constant and here $\lambda$ increases monotonically from zero 
to unity over a time long compared to a Kepler period. We denote the
quasi-bound states of $H_{1}$ when $\lambda=1$ by $\ket{n\,F}$. If initially 
the atom is in the state $\ket{\nze\,0}$ there is no classical ionisation  
provided $F<\Fcrit$, where, in scaled units, 
$\Fcrit=2^{10}/(3\pi)^{4}\simeq 0.1298$,  Richards (1987),
and complete ionisation if $F > \Fcrit$. In quantum mechanics tunnelling 
decreases these thresholds by an amount that depends upon the interaction
time and the initial principal quantum number.  For an excited 
$1d$ hydrogen atom initially in state $\ket{n\,F}$ 
the probability of remaining bound at time $t$ can be deduced from the 
relations derived in Richards (1987) and  behaves approximately as,
\[
P_{b}(n,t) \sim \exp(-\Gamma t), \quad n^{3}\Gamma\simeq \frac{1}{2\pi}
\exp\left( -2.58nc(n)\left( \frac{\Fcrit-F}{\Fcrit }\right)\right),
\quad (F < \Fcrit)
\]
where for $n > 20$, $c(n) \simeq 1/(1+1.65/n +173.1/n^{2} - 249.5/n^{3})$
is derived by a numerical fit to the theory.
This probability approaches a step function as $n\to\infty$. For 
$t=2\pi n^{3}$, that is one Kepler period, the probability $P_{b}$ 
decreases from 0.9 to 0.1 as $F$ increases over an interval $\gamma\Fcrit$
where $\gamma=0.2$, 0.1, 0.04 and 0.02 for $n=5$, 10, 30 and 50, 
respectively.

For $F<\Fcrit$ the state 
$\ket{\nze\,F}$ can be approximated by a linear combination of the field free
states; the matrix elements $\wave{\nze\,F}{n\,0}$ provide some idea 
of how many
unperturbed states are required to accurately describe the wave function
in the presence of a strong field. This matrix element is estimated by
Richards {\em et al} (1989) where it is shown to be significant for
$\nze < n < m$, where $m\simeq \nze (2F)^{-1/4}\,(=1.5\nze$ for 
$F=0.1)$. Thus any realistic approximation using a basis of unperturbed 
states requires about $2\nze$ states in a $1d$ system and about 
$\nze^{2}$ states for each of the $n$ values of the 
azimuthal quantum number, $m$. The method used by Robicheaux {\em et al} (2002)
avoids these problems, but as $\nze$ increases the number of grid points 
increases and the computational time increases commensurately.

The theory presented here minimises coupling between basis states 
by describing 
the motion in a basis that diagonalises the static-field, or Stark, Hamiltonian
\drlabel{eq:2-04}
\begin{equation}
H_{S}=\frac{1}{2\mu} \bp^{2} - \frac{e^{2}}{r} + Fz, \quad
F={\rm constant}\;(\geq 0).
\label{eq:2-04}
\end{equation}
If $\ket{\bn\,F}$ is a bound eigenstate of $H_{S}$ with energy 
$E_{\bn}(F)$ then a basis that may be used to describe the bound 
motion of the time-dependent Hamiltonian is obtained simply by 
replacing the constant $F$ by $F(t)$; in this basis coupling 
between states is caused only by the rate of change $F(t)$  not by its
magnitude. For the examples of interest here it will be seen that
this coupling is, in scaled units, $O(\omo \Fmu)$, which is typically an
order of magnitude smaller than the coupling between the unperturbed states. 
This method was first used in the present context
by Richards (1987) and Richards {\em et al} (1989), where it was applied to the
one-dimensional hydrogen atom and shown to explain important features of the
three-dimensional experimental results. 

Because this approximation uses a bound-state basis the continuum needs to
be introduced as an extra approximation, described later. In addition,
the eigenfunctions $\wave{\br}{\bn\,F}$ are not conveniently represented 
by simple functions, consequently
approximations to these are necessary. Finally, since canonical transformations
are easier to handle than their corresponding quantal unitary 
transformations, it
is easier to develop this approximation using classical dynamics and to 
quantise the resulting Hamiltonian, rather than tackle the quantum mechanics
directly.

\subsection*{The classical Stark effect}
The first goal is to find a suitable approximation to the generating function, 
$S(\bI,\br,F)$, for the canonical transformation to the angle-action variables 
of $H_{S}$; we also need expressions for $H_{S}$ and $\cd S/\cd F$ in terms of 
these variables. This is a relatively routine, but complicated, calculation 
because it is necessary to expand to high orders in $F$. The main 
result of these calculations is the adiabatic Hamiltonian, defined by 
equation~\ref{eq:2-13} below, which forms the basis of further approximations.
 
The theory starts with the Coulomb-Stark Hamiltonian, equation~\ref{eq:2-04}, 
in which the force on the electron is static and in the negative $z$-direction.
This Hamiltonian is separable in the parabolic coordinates; following 
Born (1960, section~35) we use the coordinates
\[
x=\xi\eta\cos\phi, \quad y=\xi\eta\sin\phi, \quad z=\frac12
(\xi^{2}-\eta^{2}), \quad \xi\geq 0,\;\eta\geq 0,
\]
sometimes named squared parabolic coordinates, giving
\drlabel{eq:2-05}
\begin{equation}
H_{S}=\frac{1}{2\mu(\xi^{2}+\eta^{2})}
\left( p^{2}_{\xi} +p^{2}_{\eta} +
\frac{\xi^{2}+\eta^{2}}{\xi^{2}\eta^{2}}p^{2}_{\phi} - 4\mu e^{2}\right)
+ \frac12 F(\xi^{2}-\eta^{2})=E.
\label{eq:2-05}
\end{equation}  
In the following we assume $F \geq 0$ and $E < 0$.
The Hamilton-Jacobi equation is
\drlabel{eq:2-06}
\begin{equation}
\left( \pad{S}{\xi}\right)^{2} + \left(\pad{S}{\eta} \right)^{2}
+\left( \frac{1}{\xi^{2}} + \frac{1}{\eta^{2}} \right)
\left( \pad{S}{\phi}\right)^{2} +
\mu F(\xi^{4}-\eta^{4})-4\mu e^{2}=2\mu E(\xi^{2}+\eta^{2}),
\label{eq:2-06}
\end{equation}  
the general solution of which defines
the generating function $S(\bI,\br,F)$ for the canonical transformation to 
the required angle-action variables.  This equation is 
separable so $S=S_{1}(\xi)+S_{2}(\eta) + \Ism\phi$ where
\drlabel{eq:2-07a/b}
\begin{eqnarray}
S_{1}(\xi)  & = & \int \frac{d\xi}{\xi} \left( -\mu F\xi^{6}-2\mu 
|E|\xi^{4}+2\alpha_{1}\mu e^{2}\xi^{2} -\Ism^{2}  \right)^{1/2},
\label{eq:2-07a} \\
S_{2}(\eta) & = & \int \frac{d\eta}{\eta} \left( \mu F\eta^{6}-2\mu 
|E|\eta^{4}+2\alpha_{2}\mu e^{2}\eta^{2} -\Ism^{2}  \right)^{1/2},
\label{eq:2-07b}
\end{eqnarray} 
where $\alpha_{1}$ and $\alpha_{2}$ are the dimensionless separation 
constants that satisfy $\alpha_{1}+\alpha_{2} = 2$, with  $\alpha_{1}>0$
and $\alpha_{2}>0$. Motion in the $\xi$-direction can be bound or unbound 
whereas motion in the $\eta$-direction is always bound.
The bound motion is restricted to the regions $0\leq \xi_{1}\leq \xi \leq 
\xi_{2}$ and
$0\leq \eta_{1} \leq \eta \leq \eta_{2}$ where $\xi_{1}$ and $\eta_{1}$ 
are zero only if $\Ism=0$.

The action variables are defined by the integrals
\drlabel{eq:2-08a/b}
\begin{eqnarray}
\Ion & = & \frac{1}{\pi} \int_{\xi_{1}}^{\xi_{2}} \frac{d\xi}{\xi} 
\left( -\mu F\xi^{6}-2\mu |E|\xi^{4}+2\alpha_{1}\mu e^{2}\xi^{2} -\Ism^{2}  
\right)^{1/2},
\label{eq:2-08a}\\
\Itw & = & \frac{1}{\pi}\int_{\eta_{1}}^{\eta_{2}}\frac{d\eta}{\eta} 
\left( \mu F\eta^{6}-2\mu |E|\eta^{4}+2\alpha_{2}\mu e^{2}\eta^{2} -
\Ism^{2}  \right)^{1/2},
\label{eq:2-08b}
\end{eqnarray} 
and satisfy the relation $\Ion+\Itw+|\Ism|=\In$ and 
$0 \leq I_{1,\,2} \leq \In-|\Ism|$. They are related to the usual quantum 
numbers $n_{1}$ and $n_{2}$ by 
\[
I_{k}=(n_{k}+1/2)\hbar \quad {\rm with} \quad  n_{1}+n_{2}+|m|+1=n, 
\quad 0\leq n_{k}\leq n-|m|-1.
\]
These equations relate $(\Ion,\Itw)$ to $(E,\alpha_{1})$ and 
may be
inverted to give $E$ and $\alpha_{1}$ in terms of $(\Ion,\Itw)$. However,
for $F\neq 0$ the integrals cannot be evaluated in closed form.
One method of inverting these equations is to invert the series obtained by
expanding as a power series
in $F$. A method of performing these calculations is outlined in the 
appendix; the resulting algebra is complicated and performed using Maple.  
For reasons that will soon become apparent, we have computed these series 
to $O(F^{17})$, but here quote lower
order expansions. The resulting perturbation series for the energy is
\drlabel{eq:2-09}
\begin{equation}
E(\bI)=-\frac{\mu e^{4}}{2\In^{2}} + \sum_{k=1}^{\infty}E_{k}(\bI)F^{k},
\quad (\In=\Ion+\Itw+|\Ism|),
\label{eq:2-09}
\end{equation}
where 
\begin{eqnarray*}
E_{1}(\bI) &=& - \frac32 \frac{\In\Ie}{\mu e^{2}}F , \quad \quad 
E_{2}(\bI)=- \frac{\In^{4}F^{2}}{16\mu^{3}e^{8}} 
\left( 17\In^{2}-3\Ie^{2} - 9\Ism^{2} \right),  \\
E_{3}(\bI)&=&-\frac{3\In^{7}\Ie F^{3}}{32\mu^{5}e^{14}}  
\left( 23\In^{2}-\Ie^{2}+11\Ism^{2} \right), \\
E_{4}(\bI)&=& -\frac{3\In^{10}F^{4}}{1024\mu^{7}e^{20}}  
\left( 1829\In^{4} -1134\Ism^{2}\In^{2}- 183\Ism^{4}
+(602\In^{2}-378\Ism^{2})\Ie^{2} + 49\Ie^{4}\right),  \\
E_{5}(\bI)&=& -\frac{3\In^{13}\Ie F^{5}}{1024\mu^{9}e^{26}}  
\left( 10563\In^{4} + 772\In^{2}\Ism^{2}+
725\Ism^{4} +(98\In^{2}+220\Ism^{2})\Ie^{2} -21\Ie^{4} \right), \\
E_{6}(\bI)&=& -\frac{\In^{16}F^{6}}{8192\mu^{11}e^{32}} \left\{ \vsp{10}
547262\In^{6}-429903\In^{4}\Ism^{2}-16200\In^{2}\Ism^{4}-6951\Ism^{6} 
\right.   \\ 
&& \left. +\left( 685152\In^{4}-25470\Ism^{2}\In^{2} - 36450\Ism^{4} \right)
\Ie^{2} +\left( 390\In^{2}+ 765\Ism^{2}\right)\Ie^{4} - 372\Ie^{6}
\vsp{10}\right\},\\
E_{7}(\bI)&=& -\frac{3\In^{19}\Ie F^{7}}{32768\mu^{13}e^{38}} \left\{ \vsp{9}
7071885\In^{6}-1530561\In^{4}\Ism^{2}+94915\In^{2}\Ism^{4}+55937\Ism^{6} 
\right. \\
&&+\left. \vsp{9} \left( 1502283\In^{4}+21410\In^{2}\Ism^{2}+
66115\Ism^{4}\right)
\Ie^{2}  +\left( 1947\In^{2}-6321\Ism^{2}\right)\Ie^{4} + 957\Ie^{6} \right\},
\end{eqnarray*}
and $\Ie=\Itw-\Ion$; this last action
variable is related to the electric quantum number, $\Ie=n_{e}\hbar$, though
in some text, for instance Bethe and Salpeter (1957), the electric quantum 
number is defined to be $n_{1}-n_{2}$, because coordinates are chosen so
the force on the electron
is in positive $z$-direction; the value of $\Ie$ is the
projection of the Runge-Lenze vector along $Oz$. Up to $O(F^{5})$ the above 
expansion agrees with the series given in Damburg and Kolosov (1983, page~45), 
see also Silverstone (1978),
as $\hbar\to 0$. Note that the odd components $E_{2k+1}(\bI)$ have a term
linear in $\Ie$; it will be shown that these components determine
the resonance position. The separation constant is, to $O(F^{3})$,
\drlabel{eq:2-09a}
\begin{eqnarray}
\alpha_{1}  & = &\frac{2\Ion+\Ism}{\In} + 
\frac14 \frac{\In^{2}F}{\mu^{2}e^{6}} (3\In^{2}-3\Ie^{2} -\Ism^{2})
 - \frac18 \frac{\In^{5}\Ie F^{2}}{\mu^{4}e^{12}} \left( \In^{2}-\Ie^{2} - 
6\Ism^{2}\right) \label{eq:2-09a}\\
&&+\frac{1}{128} \frac{\In^{8}F^{3}}{\mu^{6}e^{18}} 
\left\{ \vsp{8} \left(171\In^{2}-15\Ie^{2}  \right) 
\left( \In^{2}-\Ie^{2}  \right) 
-\Ism^{2} \left( 82\In^{2}+150\Ie^{2}+25\Ism^{2} \right) \vsp{12} \right\}
 +O(F^{4}) \nonumber
\end{eqnarray}  
with $\alpha_{2}=2-\alpha_{1}$.

The angle-variables corresponding to the action variables defined in 
equations~\ref{eq:2-08a} and~\ref{eq:2-08b} are defined by 
$\theta_{k}=\cd S/\cd I_{k}$, $k=1,\,2$. It is shown in the appendix 
that the two angle variables can be expressed in terms of relations
\drlabel{eq:2-10a}
\begin{equation}
\thetaon=\psi + P_{1}(\psi) + Q_{1}(\chi), \quad
\thetatw=\chi + P_{2}(\psi) + Q_{2}(\chi), \quad
\label{eq:2-10a}
\end{equation}
where the two auxiliary angles, $(\psi,\chi)$ are defined by the relations
\drlabel{eq:2-10c}
\begin{equation}
\xi^{2}=\frac12\left( \xi_{2}^{2}+\xi_{1}^{2}\right) - 
\frac12\left( \xi_{2}^{2}-\xi_{1}^{2} \right)\cos\psi, \quad
\eta^{2}=\frac12\left( \eta_{2}^{2}+\eta_{1}^{2}\right) - 
\frac12\left( \eta_{2}^{2}-\eta_{1}^{2} \right)\cos\chi,
\label{eq:2-10c}
\end{equation}
and where $(P_{k}(x),Q_{k}(x))$ are odd, $2\pi$-periodic functions with zero
mean value; expressions for these, accurate up to $O(F)$, are given
in the appendix, equation~\ref{eq:app-ang05}. However, for reasons 
discussed below, we require these 
functions only in the limit $F=0$ and then we have
\drlabel{eq:2-10}
\begin{equation} 
\thetaon = \psi - \sigma_{1}\sin\psi - \sigma_{2}\sin\chi, \quad
\thetatw = \chi - \sigma_{1}\sin\psi - \sigma_{2}\sin\chi, 
\quad (F=0), \label{eq:2-10}
\end{equation} 
where $\sigma_{k}=\sqrt{I_{k}(I_{k}+\Ism)}/\In$, $k=1,\,2$.

\subsection*{Dynamic Stark effect}
When the field amplitude $F$ varies with time the function, 
$S(\bI,\br,F(t))$, generates a time-dependent canonical transformation and
the Hamiltonian becomes
\drlabel{eq:2-11}
\begin{equation}
K= E(\bI,F(t)) + \pad{S}{F}\frac{dF}{dt}.
\label{eq:2-11}
\end{equation}  
The first term of this is the Stark Hamiltonian, equation~\ref{eq:2-09};
the second term is more difficult to find, but it is important
because only this term mixes states. It is shown in the appendix that
the function $\cd S/\cd F$ can be 
expressed as a Fourier series of the following form
\drlabel{eq:2-10b}
\begin{equation}
\pad{S}{F}=\sum_{k=1}^{\infty} A_{k}\sin k\psi + B_{k}\sin k\chi
\label{eq:2-10b}
\end{equation}
where the angles $(\psi,\chi)$ are defined in equation~\ref{eq:2-10c}
and where the coefficients $(A_{k},B_{k})$ are functions of the action 
variables and $F$. It is important to note that there is no 
term independent of both $\psi$ and $\chi$. In our applications
$F(t)=\lambda(t)(\Fs + \Fmu\cos\Omega t)$ and 
$\dot{\lambda}/\lambda \ll \Omega$, so
$\dot{F}\simeq -\lambda\Omega\Fmu\sin\Omega t$ and since,
in scaled units $\Omega$, $\Fs$ and $\Fmu$ are numerically similar and small, 
a second order approximation is obtained by evaluating the derivative 
$\cd S/\cd F$ at $F=0$, which considerably simplifies the analysis: 
the following result is derived in the appendix,
\[
\pad{S}{F} =  \frac{\In^{4}}{2\mu^{2}e^{6}}G, \quad (F=0),
\]
where
\[
G =  (3\Itw+\Ion+2\Ism)\sigma_{1}\sin\psi -
(3\Ion+\Itw+2\Ism)\sigma_{2}\sin\chi -
\frac{\In}{2}\left( \sigma_{1}^{2}\sin 2\psi-
\sigma_{2}^{2}\sin 2\chi\right).
\]
It is more convenient to use new angle-action variables, 
\drlabel{eq:2-12}
\begin{equation}
\begin{array}{ll}
\In =\Ion+\Itw+\Ism, \quad &  \thetaon=\phin-\phie, \\
\Ie=\Itw-\Ion, \quad  & \thetatw=\phin+\phie,\\
\Ism=\Jm, \quad & \theta_{m}=\phin +\phim,
\end{array}  
\label{eq:2-12}
\end{equation}
so, when $F=0$ equation~\ref{eq:2-10} gives 
$2\phie=\thetatw-\thetaon=\chi-\psi$. This gives an approximation we name the 
{\em adiabatic} Hamiltonian,
\drlabel{eq:2-13}
\begin{equation}
K =  E(\bI,F)+\frac{\In^{4}}{2\mu^{2}e^{6}}\frac{dF}{dt} 
G(\In,\Ion,\Itw,\psi,\chi),
\label{eq:2-13} 
\end{equation}
where $E(\bI,F)$ is the Stark energy given in equation~\ref{eq:2-09}.
The angles $\psi$ and $\chi$ are not conjugate to the action variables and to
develop further approximations it is necessary to express all quantities 
involving $\psi$ and $\chi$ in terms of $\thetaon$ and $\thetatw$, using 
equations~\ref{eq:2-10}. This is most easily achieved by expressing each
function as a multiple Fourier series,
\drlabel{eq:2-14}
\begin{equation}
\left( \begin{array}{c}
\sin k\psi\\
\sin k\chi
\end{array} \right) = 
\sum_{s_{1}=-\infty}^{\infty}\sum_{s_{2}=-\infty}^{\infty}
\left( \begin{array}{c}
\disp  S_{s_{1}\,s_{2}}^{(k)} \vspace{4pt}\\
\disp C_{s_{1}\,s_{2}}^{(k)} 
\end{array}
\right) 
\exp \left[-i\left( s_{1}\thetaon+s_{2}\thetatw \right)\right], \quad k=1,\,2,
\label{eq:2-14}
\end{equation}
where $C_{s_{1}\,s_{2}}^{(k)}=S_{s_{2}\,s_{1}}^{(k)}$ and
\[
S_{s_{1}\,s_{2}}^{(k)} = \left\{ \begin{array}{l}
\disp i \frac{k}{2s}   J_{s_{2}}(s\sigma_{2})
\left[ J_{s_{1}+k}(s\sigma_{1}) + J_{s_{1}-k}(s\sigma_{1}) \right],
\quad s=s_{1}+s_{2}\neq 0,  \vspace{6pt}\\
\disp \pm \frac{i\sigma_{2}}{4}, \quad s=0,\quad s_{1}=\pm 1, \quad 
{\rm and} \quad k=1,\\
0, \quad {\rm otherwise.}
\end{array}
\right.
\]

The adiabatic Hamiltonian~\ref{eq:2-13} is useful because the coupling
term is $O(\omo\Fmu)$ rather than $O(F)$ as in the original 
Hamiltonian, so the resulting Schr\"{o}dinger equation may be solved using a 
far smaller basis. 

Hamilton's equations in the original representation are
singular at $r=0$ and this is dealt with using a regularisation method, see
Szebehely (1967) for a general introduction and Rath and Richards (1988) for
an application to the perturbed hydrogen atom. The
equivalent singularity in the adiabatic Hamiltonian occurs when $\Ism=0$ and
$\sigma_{1}+\sigma_{2}=1$ and this also needs to be removed. For instance 
the equation for $\psi$ is $\dot{\psi}=[\psi,K]$ and the right hand side
is proportional to $1/J$ where 
$J=\In\left( 1-\sigma_{1}\cos\psi - \sigma_{2}\cos\chi\right)$, which can
be zero.
A method of avoiding numerical problems when $J$ is small is to define
a new time, $\tau$, by the equation $dt/d\tau=J$, to give the equations of
motion 
\drlabel{eq:ade1}
\begin{eqnarray}
\frac{d\Ion}{d\tau} &=& -\In\kappa\frac{dF}{dt} 
\left( \vsp{10}(1-\sigma_{2}\cos\chi)G_{\psi}+
\sigma_{1}\cos\psi\,G_{\chi} \right),
\nonumber \\
\frac{d\Itw}{d\tau} &=& -\In\kappa\frac{dF}{dt} 
\left( \vsp{10} \sigma_{2}\cos\chi\,G_{\psi}+
(1-\sigma_{1}\cos\psi)\cos\psi \,G_{\chi}\right),
\nonumber \\
\frac{d\psi}{d\tau} &=& \In\pad{K}{\Ion} +
\left(\pad{K}{\Itw}-\pad{K}{\Ion} \right)\In\sigma_{2}\cos\chi
\label{eq:ade1}\\
&& \hspace{30pt} + \kappa\frac{dF}{dt} 
\left( \frac{2\Ion+\Ism}{2\sigma_{1}\In}\sin\psi -
\frac{2\Itw+\Ism}{2\sigma_{2}\In}\sin\chi\right\}G_{\chi},
\nonumber \\
\frac{d\chi}{d\tau} &=& \In\pad{K}{\Itw} +
\left(\pad{K}{\Ion}-\pad{K}{\Itw} \right)\In\sigma_{1}\cos\psi
\nonumber \\
&& \hspace{30pt} - \kappa\frac{dF}{dt} 
\left( \frac{2\Ion+\Ism}{2\sigma_{1}\In}\sin\psi -
\frac{2\Itw+\Ism}{2\sigma_{2}\In}\sin\chi\right\}G_{\psi}, \nonumber
\end{eqnarray}
where $\kappa=\In^{4}/(2\mu^{2}e^{6})$ and $\In=\Ion+\Itw+\Ism$.
In figure~\ref{f:8} below we compare ionisation probabilities 
computed using these equations and the original Hamiltonian, but first it is
necessary to re-introduce ionisation into this approximation.

\subsection{Ionisation}
\drlabel{sec:ion}\label{sec:ion}
The adiabatic Hamiltonian, equation~\ref{eq:2-13}, does not allow 
for ionisation because
angle-action variables exist only for bound orbits. Ionisation therefore has
to be included as an extra approximation which is described here.

For static fields each classical state, or torus, labelled by the actions
$(\In,\Ie,\Ism)$, has a critical field 
$\Fcrit$ such that it exists only if $0\leq F < \Fcrit$: the approximation
to $\Fcrit$ given by Banks and Leopold (1978) is used here.
Note that if $\Ism=0$ bound orbits exist for all $F$, Howard (1995), but those
orbits that exist for large $F$ are so special that they do not affect the
current problem. Adiabatic invariance suggests that this 
behaviour persists for sufficiently slowly
varying fields, that is small scaled frequencies. The extreme values of 
$\Fcrit$ occur when the atom is aligned along the field, $\Ism=0$: in scaled 
units $\min(\Fcrit)\simeq 0.13$ for $(\Ie,\Ism)=(\In,0)$, and 
$\max(\Fcrit)\simeq 0.38$ for $(\Ie,\Ism)=(-\In,0)$. The variation 
of $\Fcrit$ with $\Ie=\Itw-\Ion$, for various values of $\Ism$, is
shown in the following figure.

\vspace{10pt}

\begin{center}
\drlabel{f:7}
\begin{drfigs}
\drsetfig{240pt}{crit-field.eps}
{The scaled classical critical field $\Fcrit$ as a function of the scaled
action $\Ie=\Itw-\Ion$, for various $\Ism$.}{f:7}
\end{drfigs}
\end{center}

\noindent
In the adiabatic limit $\In$ and $\Ie$ are almost constant, so we may 
define the time-dependent critical field $\Fcrit(t)=\Fcrit(\Ie(t),\Ism)$ 
and assume that ionisation occurs at the time when 
the actual field $F(t)$ exceeds this. If $F(t)$ changes sign, as happens if 
\mbox{$\Fmu > \Fs$}, the quantisation axis changes direction and 
the relevant critical field is given by $\Fcrit( -\Ie,-\Ism)$ 
$(F<0)$ so the combined ionisation criterion is:
\drlabel{eq:st30}
\begin{equation} \begin{array}{lcll}
F(t) & > & \quad \Fcrit^{(+)}=\Fcrit(\Ie(t),\Ism), \quad {\rm if} & 
\quad F(t) >0, \vspace{4pt}\\
F(t) & <  & -\Fcrit^{(-)}=-\Fcrit(-\Ie(t),-\Ism), \quad {\rm if} & 
\quad F(t)  <0.
\end{array}
\label{eq:st30}
\end{equation} 
This approximation is accurate when the field changes very little during one 
Kepler period, that is $\omo \ll 1$, and then this  criteria may be 
used to include ionisation in the adiabatic equations of motion, 
equations~\ref{eq:ade1}.

A useful guide to the behaviour of the system is obtained by 
putting $\In(t)$ and $\Ie(t)$ equal to their initial values. This gives
two boundaries beyond which $\Pion=1$:\\[4pt]
\begin{tabular}{clll}
I: & $\Fs > \Fm$ \quad & $\Pion(\Fs)=1$  if  &
$\Fs > \Fcrit^{+}(\Ie,\Ism) -\Fm$ \vspace{4pt}\\
II: & $\Fs <  \Fm$ \quad & $\Pion(\Fs)=1$ if   &
$\Fs <  \Fm - \Fcrit^{-}(\Ie,\Ism)$ \hspace{0.5pt} or \hspace{0.5pt}
$\Fs > \Fcrit^{+}(\Ie,\Ism) -\Fm$ 
\end{tabular}\\[2pt]
where $\Ie=\Ie(0)$ is the initial value of this action variable.

For the ionisation curves  shown in figure~\ref{f:8},
$\Ie(0)=-0.6$ and $\Ism=0$ giving $\Fcrit^{+}=0.219$ and $\Fcrit^{-}=0.142$.
Thus, since $\Fm=0.15$ the condition~II gives $\Pion(\Fs)=1$ if
$\Fs< 0.008$ or $\Fs > 0.069$: these boundaries are shown by the arrows in the 
figures and are consistent with the calculations. Similar boundary-arrows are
included in figures~\ref{f:2} and~\ref{f:5}.

Non-adiabatic dynamics affect this simple picture in two ways. First, 
they blur and slightly shift the boundaries. Second, and more important,
isolated resonances produce large changes in $\Ie(t)$ and can enhance 
ionisation at particular combinations of $(\Fs,\Fm)$, other than those
defined by I and II above. These dynamical effects produce the 
peaks labelled $j=1$-$4$ seen in figure~\ref{f:8}.

In the following figure we compare values of $\Pion$ computed
using the original Hamiltonian~\ref{eq:2-01}, the solid lines, and the 
adiabatic equations~\ref{eq:ade1}, the dashed lines.
Here $\omo=0.0528$ ($\nze=35$),
$\Itw=0.2$, $\Ism=0$ and $\Fmu=0.15$ and the arrows denote the values
of $\Fmu-\Fcrit^{-}$ and $\Fcrit^{+}-\Fmu$: in the left hand panel the 
field envelope is 4-50-4 and in the right panel it is 16-50-16.

\begin{center}
\drlabel{f:8}
\begin{drfigs}
\drsetfig{300pt}{n35-adia.eps}{Comparison of ionisation probabilities 
computed using exact dynamics (solid line) and the adiabatic 
equation~\ref{eq:ade1} (dashed line) for two envelopes. See the text
for the explanation of the arrows, which point to the borders outside of
which $\Pion=1$, in the adiabatic limit.}{f:8}
\end{drfigs}
\end{center}

\noindent
These figures show broad agreement between the two calculations, but there 
are three marked differences; consider the left hand panel.
\begin{itemize}
\item  The maxima $j=2$, 3 and 4 are at different values of $\Fs$. This
is because only terms upto $O(F^{2})$
were included in the expansion of $E(\bI)$, equation~\ref{eq:2-09}, and
is not an inherent inaccuracy of the adiabatic approximation.

\item The maximum $j=1$ is not present in the adiabatic calculation
because with the approximations used it disappears when 
$\Fs=\omo/3=0.176$ and $\Fm=\omo j'_{1k}/3$, $k=1,\,2,\cdots$, 
where $j'_{jk}$ are the positive, real roots of $J'_{j}(x)=0$; with $k=3$
this gives $\Fm=0.150$. For the full Hamiltonian this resonance disappears
at $\Fs\simeq 0.0161$, $\Fm\simeq 0.147$.

\item The main difference is the shift in the shoulder near $\Fs=0.01$.
\end{itemize}

\noindent
Apart from these differences the agreement between the two approximations 
is good. The same remarks apply to the right hand panel but now we 
see that the new maxima
at $\Fs\simeq 0.039$ and some of the structure at $\Fs\simeq 0.07$ are 
reproduced in the adiabatic approximation. The adiabatic approximation also 
has a local minimum at $\Fs=0.0025$,
not present in the `exact' probabilities: however, similar behaviour is seen
in the exact results for other parameters values, see for instance the right
hand panel of figure~\ref{f:5}.

These results, and other comparisons that cannot be shown here, suggest
that the adiabatic Hamiltonian provides a good approximation to the 
true dynamics. This is important because, for the principal quantum
numbers used in current experiments, the numerical solution of the
Schr\"{o}dinger equation derived from this Hamiltonian is a feasible 
computational task, unlike that derived from the exact Hamiltonian using 
either an unperturbed or a static-Stark basis. In the quantal
approximation ionisation is included by adding a complex part to the 
energies, which may be computed semiclassically, as in
Leopold and Richards (1991) and Sauer {\em et al} (1992).

\subsection{Averaged equations of motion}
The adiabatic Hamiltonian~\ref{eq:2-13} can be simplified further by noting 
that the two natural frequencies of the motion are quite different,
\[
\omegae=\pad{K}{\Ie}=-\frac{3\In F(t)}{2\mu e^{2}}+O(F^{2}), \quad
\omega_{n}=\pad{K}{\In}=\frac{\mu e^{4}}{\In^{3}} -
\frac{3\Ie F(t)}{2\mu e^{2}}+ O(F^{2}),
\]
so $|\omegae|\ll \omega_{n}\simeq \omega_{K}$; this means that the
orbital elements of the Kepler-ellipse change relatively little during
one Kepler period. Hence the first averaged
approximation is obtained by averaging over $\phin$, which is most easily
achieved by ignoring all terms containing $\phin=(\thetaon+\thetatw)/2$ 
in the Fourier series~\ref{eq:2-14}. This gives
\[ 
\mean{ \sin\psi }=-\frac12 \sigma_{2}\sin 2\phie, \;
\mean{ \sin\chi }=\frac12 \sigma_{1}\sin 2\phie
\; {\rm and} \;\mean{ \sin k\psi} = \mean{ \sin k\chi }=0,\; k\geq 2.
\]
Substituting these mean values into the adiabatic Hamiltonian~\ref{eq:2-13} 
gives the mean motion Hamiltonian
\drlabel{eq:st15}
\begin{equation}
\Kmean = E(\Ie,F(t))
-\frac14 \frac{\In^{3}}{\mu^{2}e^{6}}\frac{dF}{dt} 
A(\Ie)B(\Ie) \sin 2\phie \label{eq:st15}
\end{equation}
where 
\[
A(\Ie)^{2}= (\In+\Ism)^{2}-\Ie^{2}, \quad B(\Ie)^{2}= (\In-\Ism)^{2}-\Ie^{2}.
\]
In quantum mechanics this approximation corresponds  to ignoring all 
transitions between states of different $n$-manifolds of the adiabatic basis.

Numerical integration of the equations of motion derived from \ref{eq:st15}, 
is not straightforward because of the 
square root singularity of $B(\Ie)$ at $\Ie=\pm(\In-\Ism)$. It is best 
accomplished by introducing the three-dimensional vector
\[
\bZ=(B(\Ie)\cos 2\phie,B(\Ie)\sin 2\phie,\Ie),
\]
the components of which satisfy the commutation relations
\(
[Z_{i},Z_{j}]=2\epsilon_{ijk}Z_{k}
\)
and $|\bZ|^{2}=$constant so the vector $\bZ$ moves on the surface of a sphere.
The equations of motion, $\dot{Z}_{k}=[Z_{k},H]$, become
\drlabel{eq:st16}
\begin{eqnarray}
\frac{dZ_{1}}{dt} & = & -2Z_{2}\pad{E}{Z_{3}} -
\kappa   \left\{ \vsp{9}
Z_{3}A(Z_{3}) -Z_{2}^{2}A'(Z_{3}) \right\}\frac{dF}{dt},
\nonumber \\
\frac{dZ_{2}}{dt}  & = & 2Z_{1}\pad{E}{Z_{3}} -
\kappa Z_{1}Z_{2}A'(Z_{3})\frac{dF}{dt}, \quad
\frac{dZ_{3}}{dt}  =  \kappa  Z_{1}A(Z_{3})\frac{dF}{dt},
\label{eq:st16}
\end{eqnarray} 
where $\kappa=\In^{3}/(2\mu^{2}e^{6})$. These equations 
are trivially solved numerically.

\subsection{The Resonance Hamiltonian}
\drlabel{sec:resham}\label{sec:resham}
The mean motion Hamiltonian~\ref{eq:st15} needs to be rearranged in order to
extract a clear picture of the dynamics. Observe that the odd terms
in the series for the Stark energy, equation~\ref{eq:2-09}, 
contain components that are linear in $\Ie$ and that these terms produce a 
slow secular change in $\phie$ which physically corresponds to
a rotation of the angular momentum vector about the Runge-lenze vector. 
We shall see that the mean part
of this motion determines the position of the resonances and the 
oscillatory part causes these resonances to disappear at 
certain field ratios. 

Denoting these linear terms by $E_{L}(\Ie,F)$ and on setting $\mu=e=1$, we
have, to $O(F^{5})$
\[
E_{L}=-\frac32 \In\Ie F \left( 1 + 
\frac{\In^{6}}{16}\left( 23\In^{2}+11\Ism^{2}\right) F^{2} +
\frac{\In^{12}}{512}( 10563\In^{6}+772\In^{2}\Ism^{2}+725\Ism^{4})F^{4}
+\cdots\right).
\]
The remaining part of the Hamiltonian, $\dot{F}\cd S/\cd F$, has a different
form, equation~\ref{eq:2-10b}, and does not give rise to factors like
$\Ie F^{k}$; this is important for the following analysis.

Using $E_{L}$ only in Hamilton's equations we find that
$\phie(t)=\phie(0)-3\In g(t)/2$ where 
\drlabel{eq:rh-01}
\begin{equation}
g(t)=\int_{0}^{t}dt\,\left[ F+  
\frac{\In^{6}}{16}\left( 23\In^{2}+11\Ism^{2}\right)F^{3}+
\frac{\In^{12}}{512}(10563\In^{4}+772\In^{2}\Ism^{2}+725\Ism^{4})F^{5}
+\cdots\right].
\label{eq:rh-01}
\end{equation}
The series in this integral has a finite radius of convergence, $F_{rc}(\Ism)$,
so it is important that $\max(F) < F_{rc}$. We have computed this series to 
$O(F^{17})$, and used these nine terms to estimate $F_{rc}$. By 
extrapolating
the ratios of coefficients using Richardson's extrapolation we estimate
that $F_{rc}\sim 0.17$, 0.19 and 0.21 for $\Ism=0$, 0.8 and 1, respectively.
Using Pad\'{e} approximants we obtain $F_{rc}\sim 0.18$, 0.20 and 0.22,
respectively. This provides a rough guide to the range of fields for
which the following theory is valid.

Since the field amplitude, $F(t)$, is periodic in $t$, the function $g(t)$ 
can be written in the form $g(t)=\gb t+\gt(t)$, where $\gt(t)$ is periodic
with zero mean and $\gb$ is the mean of~$E_{L}$ over a field period.
With $F(t)=\Fs+\Fm\cos\Omega t$ this becomes
\drlabel{eq:rh-02}
\begin{eqnarray}
\gb &=& \Fs + \frac{\In^{6}}{16}\left( 23\In^{2}+11\Ism^{2}\right)\Fs 
\left( \Fs^{2}+\frac32\Fm^{2}\right) \nonumber\\
 && \hspace{5pt}
+\frac{\In^{12}}{512}(10563\In^{4}+772\In^{2}\Ism^{2}+725\Ism^{4})\Fs
\left( \Fs^{4}+5\Fs^{2}\Fm^{2}+\frac{15}{8}\Fm^{4}\right)+\cdots\,.
\label{eq:rh-02}
\end{eqnarray}
The periodic function $\gt(t)$ can be expressed as the Fourier series
\[
\gt(t)=\frac{\Fm}{\Omega}\sum_{k=1}^{\infty}\gt_{k}\sin k\Omega t
\]
where \drlabel{eq:rh02a}
\begin{eqnarray}
\gt_{1} &=& 1+\frac{3\In^{6}}{16}
\left( 11\Ism^{2}+23\In^{2}\right)
\left( \Fs^{2}+\frac14\Fm^{2}\right)+O(F^{4}),  \nonumber \\
\gt_{2} &=& \frac{3\In^{6}}{64}\Fs\Fm
\left( 11\Ism^{2}+23\In^{2}\right)  +O(F^{4}), \label{eq:rh-02a}\\
\gt_{3} &=& \frac{\In^{2}}{192}\Fm^{2}
\left( 11\Ism^{2}+23\In^{2}\right)  +O(F^{4}), \quad \gt_{k}=O(F^{4}),
\quad k\geq 4. \nonumber
\end{eqnarray}
The dominant harmonic is $\gt_{1}$: both $\gt_{2}$ and $\gt_{3}$ are
$0(F^{2})$ and all higher harmonics are $O(F^{4})$ and may be neglected.

Resonances in the dynamics occur when the angular frequency, 
$3\In \gb/2$, resonates
with the field frequency: the magnitude of the effect of any resonance 
depends upon the periodic component of $g(t)$, and principally upon $\gt_{1}$.

In order to see this we change to a moving reference frame, in which
$\phie(t)$ is approximately stationary, by defining a new angle 
$\psie=\phie + 3\In g(t)/2$ using the generating function
\(
F_{2}(p,\phie)=\left( \phie + 3\In  g(t)/2 \right) p, 
\)
where $(\psie,p)$ are the new conjugate variables. Since $g(t)$ is, by
definition, independent of $\Ie$ we have $p=\Ie$. In these variables the 
Hamiltonian~\ref{eq:st15} becomes
\drlabel{eq:st17}
\begin{equation}
\Kbb=\left\{ \vsp{10} E(\Ie,F(t))-E_{L}(\Ie,F(t)) \right\}
-\frac14\frac{dF}{dt} \In^{3}
A(\Ie)B(\Ie)\sin\left( 2\psie - 3\In g(t)\right).
\label{eq:st17}
\end{equation}  
No further approximation has been made in deriving this Hamiltonian from 
$\Kmean$ defined in equation~\ref{eq:st15}. By definition the curly brackets 
contains terms quadratic and
higher in $\Ie$ which are independent of $\psie$; the leading term is
\[
E(\Ie,F)-E_{L}(\Ie,F)=\frac{3}{16}\In^{4}\Ie^{2}F(t)^{2},
\]
and since $\dot{F}=\Fm\Omega\sin\Omega t$ we see that the terms of
$\Kbb$ are $O(F^{2})$ and $O(F\Omega)$, and since $\Omega \sim F \sim 0.1$ (in 
scaled units) the mean motion of $(\psie,\Ie)$ is slow by comparison with
the field oscillations: hence the relatively high 
frequency oscillations of $E-E_{L}$ do not qualitatively affect the 
motion --- a fact that has been confirmed numerically --- and hence we may
replace $E-E_{L}$ by its mean over a field period. Retaining only the 
dominant quadratic term gives
\[
\Kbb=\frac{3}{16}\left( \Fs^{2}+\frac12\Fm^{2}\right)\In^{4}\Ie^{2} +
\frac14 \In^{3}\Fm\Omega A(\Ie)B(\Ie)\sin\Omega t 
\sin\left( 2\psie - 3\In g(t)\right).
\]
The second term is, for most values of
$\Omega$, $\Fmu$ and $\Fs$, an oscillatory function of time with small mean 
value: in these circumstances it has little effect and may be ignored so
$\Kbb \sim \Ie^{2}$, giving $\Ie(t) \sim \,$constant with $\psie$ 
approximately
proportional to $t$. However, for any given $(\Omega,\Fmu)$ there are 
particular resonant values of $\Fs$ for which the long-time 
average of the second
term is proportional to $\sin 2\psie$ and then the nature of the resonant
motion is qualitatively different. Near these values of $\Fs$, $\Ie(t)$ can 
vary over a large portion of its accessible range and in some circumstances
this leads to enhanced ionisation.

The function $\gt(t)$ is periodic and odd so we may write
\drlabel{eq:st18a}
\begin{equation}
\sin \left( 2\psie -3\In g(t)\right) =
\sum_{k=-\infty}^{\infty} \sJ_{k}
\sin \left( 2\psie -\nu_{k}t+k\pi\right), \quad \nu_{k}=3\In\gb-k\Omega, 
\label{eq:st18a}
\end{equation}
where the Fourier coefficients, $\sJ_{k}$, depend upon $\gt_{s}$, 
$s\geq 1$. The coefficient $\sJ_{k}$ is dominated by $\gt_{1}=1+O(F^{2})$, but
$\gt_{2}$ and $\gt_{3}$ are also $O(F^{2})$, and to this order 
\drlabel{eq:st18b}
\begin{eqnarray}
\sJ_{k} &=& J_{k}(z_{1})J_{0}(z_{2})J_{0}(z_{3}) +
J_{0}(z_{2}) \sum_{s=1}^{\infty} J_{s}(z_{3}) 
\left[ \vsp{10}J_{k-3s}(z_{1}) + (-1)^{s}J_{k+3s}(z_{1}) \right] \nonumber\\
 && \hspace{20pt}  +J_{0}(z_{3}) \sum_{s=1}^{\infty} J_{s}(z_{2}) 
\left[ \vsp{10} J_{k-2s}(z_{1}) + (-1)^{s}J_{k+2s}(z_{1}) \right] \nonumber \\
&& \hspace{20pt} +\sum_{s=1}^{\infty}J_{s}(z_{2}) \sum_{r=1}^{\infty} 
J_{r}(z_{3}) \left[ \vsp{10}
J_{k-2s-3r}(z_{1}) + (-1)^{r}J_{k-2s+3r}(z_{1}) \right. \nonumber \\
&& \hspace{80pt} \left. 
+(-1)^{s}J_{k+2s-3r}(z_{1}) + (-1)^{s+r}J_{k+2s+3r}(z_{1}) \vsp{10}\right],
\label{eq:st18b}
\end{eqnarray}
where $z_{k}=3\gt_{k}\In\Fm/\Omega$.

Using equation~\ref{eq:st18a} the mean-motion Hamiltonian becomes,
\drlabel{eq:st18}
\begin{equation}
\Kbb =\frac{3\In^{4}}{16}
\left( \Fs^{2}+\frac12 \Fmu^{2}\right) \Ie^{2} +
\frac14 \In^{3}\Omega\Fmu
A(\Ie)B(\Ie)\sum_{k=-\infty}^{\infty} \sJt_{k}
\cos(2\psie -\nu_{k}t +k\pi),
\label{eq:st18}
\end{equation}  
where the functions $A(\Ie)$ and $B(\Ie)$ are defined after 
equation~\ref{eq:st15} and $\sJt_{j}=(\sJ_{j-1}-\sJ_{j+1})/2$, so that, to 
the lowest order, $\sJt_{j}=J'_{j}(3\Fm/\omo)$.

If $|\nu_{j}|$ is small the $j$th term of the sum changes
more slowly than all other terms, which may therefore be averaged over to give
the {\em resonance} Hamiltonian, 
\[
\Kr=\frac{3\In^{4}}{16} 
\left( \Fs^{2}+\frac12 \Fmu^{2} \right) 
\Ie^{2} + \frac14\In^{3} \Omega\Fmu A(\Ie)B(\Ie) \sJt_{j}
\cos (2\psi_{e}-\nu_{j}t+j\pi).
\]
We {\em define} the $j$th dynamical resonance to be at the static field, 
$\Fs^{(j)}$, where $\nu_{j}=0$, that is where coupling 
between $\Ie$-states is largest: the equation for $\Fs^{(j)}$ is
\drlabel{eq:st19}
\begin{equation}
3\gb(\Fs,\Fm,\Ism)\In= j\Omega \quad {\rm or\;to\;lowest\;order,\;in\;scaled\;
units,}\quad 
\Fs^{(j)}\simeq \frac13 j\omo, \quad j=1,\,2,\cdots\,.
\label{eq:st19}
\end{equation}
We show below that near this value of $\Fs$ ionisation may be
enhanced, but  in section~\ref{sec:envel}, it is shown that there is 
no clearly defined, precise relation between $\Fs^{(j)}$ and $\sF_{s}^{(j)}$,
the position the maximum 
in the ionisation probability seen in figures~\ref{f:1} and~\ref{f:2}; the
two fields are close but the difference can be larger than the resonance 
width, see section~\ref{sec:res-ion-II}, in particuar table~\ref{t:10}.

The position of the $j$th resonance, $\Fs^{(j)}$, is, to a first approximation,
independent of the substate quantum numbers; if one substate is ionised 
by this resonance others will be similarly affected so the effect of the
resonance is not significantly changed by an average over substates. 

With the translation $\theta=\psi_{e}-\nu_{j}t/2 +j\pi/2$, the 
resonance Hamiltonian may be converted to the conservative system,
\drlabel{eq:st20a}
\begin{equation}
\Kbr =\frac{3\In^{4}}{16}  \left( \Fs^{2}+\frac12\Fmu^{2}\right) 
\left(\Ie - \alpha_{j} \right)^{2} +
\frac14  \In^{3}\Omega\Fmu A(\Ie)B(\Ie) \sJt_{j}
\cos 2\theta 
\label{eq:st20a}
\end{equation} 
where
\[
\alpha_{j}= \frac{4(3\gb-j\omo)}{3\left( \Fs^{2}+\frac12\Fm^{2}\right)}\simeq 
\frac{4(3\Fs-j\omo)}{3(\Fs^{2}+\frac12\Fmu^{2})} 
\quad ({\rm in\;scaled\;units}) .
\]
The Hamiltonian $\Kbr$ shows that the $j$th resonance disappears when 
$\sJt_{j}=0$: to the lowest order this gives, in scaled units
\drlabel{eq:st20b}
\begin{equation}
\Fs^{(j)}=\frac13 j\omo \quad {\rm and} \quad \Fm^{(j,k)}=\frac13 j'_{jk}\omo,
\quad k=1,\,2,\cdots
\label{eq:st20b}
\end{equation}
where $j'_{j\hsp{0.25}k}$ is the $k$th positive zero of $J'_{j}(x)$. 
This critical value of $\Fm$ 
was first derived in a linear quantal approximation, Galvez {\em et al} 
(2000), and later by Oks and Uzer (2000) using a Floquet approximation and 
by Ostrovsky and Horsdal-Pedersen (2003) using a linear approximation. 
Recent experiments, Schultz (2003) and Shultz {\em et al} (2004),
and comparisons with classical calculations suggest that 
this estimate can be inaccurate by up to 10\%. Later we show that the
present theory can be used to improve upon these estimates.

The derivation of the resonance position~\ref{eq:st19} and the resonance
Hamiltonian~\ref{eq:st20a} involves a series of approximations. Before
proceeding it is useful 
to list these in order to estimate the effect of ignored terms.

\begin{enumerate}
\item \label{note:01}We have used the Stark angle-action variables, defined 
by $H_{S}$,
equation~\ref{eq:2-04}. For the action variables we use a series representation
in $F$. For the angle variables we use the $F=0$ limit because these variables
appear in the Hamiltonian $K$, equation~\ref{eq:2-11}, only in the term
which is $O(\omo\Fm)$.

\item \label{note:02} The term $\cd S/\cd F$, equation~\ref{eq:2-10b}, has
zero mean value when averaged over $(\psi,\chi)$ and because it is 
multiplied by the factor $\omo\Fm$ we may approximate it by its value at
$F=0$.

\item The mean motion Hamiltonian, $\Kmean$ equation~\ref{eq:st15}, is 
derived by
averaging over $\phin=(\thetaon+\thetatw)/2$ which replaces $\sin k\psi$ and
$\sin k\chi$ by Fourier series in $\sin 2\phie$. The approximations
described in points~\ref{note:01} and~\ref{note:02} truncate this series at the
first term and approximate its coefficient to second order.
\end{enumerate}

\noindent
The inclusion of higher-order terms  in the mean-motion
Hamiltonian~\ref{eq:st15} introduces corrections $O(\Omega\Fm F)$ to the
factor $\Omega\Fm A(\Ie)B(\Ie)$ and adds further terms corresponding to the
harmonics $\sin 2p\phie$, $p=2,\,3,\cdots$. Crucially this means that the 
estimate, $\nu_{j}=0$, of the $j$ resonance position is not affected by the 
approximations made: that is, the 
position of the dynamical resonance is determined solely by the
parts of the Stark Hamiltonian, $E(\bI)$, equation~\ref{eq:2-09}, which are
linear in $\Ie$.

A better estimate of $\Fs^{(j)}$ is therefore obtained using the 
series~\ref{eq:rh-02}, which has been evaluated, using computer
assisted algebra, to $O(F^{17})$. In section~\ref{sec:res-ion-II} we use this 
to obtain better estimates of $\Fs^{(j)}$
and the values of $\Fm$ at which the resonances disappear.

\section{Resonant Ionisation}
\drlabel{sec:res-ion-I}\label{sec:res-ion-I}
\subsection{Qualitative discussion}
Here we show how the dynamical resonance described in the previous
section can enhance ionisation. The connection is qualitative, but explains
many featues of the ionisation probablity.

The Hamiltonian $\Kbr$, equation~\ref{eq:st20a}, is similar to that of 
a vertical pendulum, but there are two significant differences. First,
$\Ie$ is confined to the region $|\Ie| \leq \In-\Ism$ with
natural boundaries at $\Ie=\pm(\In-\Ism)$, where $B(\Ie)=0$, see
equation~\ref{eq:st15}. Second, the
coefficient of $\cos 2\theta$ depends upon $\Ie$. The fixed
points of $\Kr$ are at the roots of $\cd\Kbr/\cd \Ie=\cd\Kbr/\cd\theta=0$, and
for this analysis it is convenient to replace $\sJt_{j}$ by $|\sJt_{j}|$; 
when $\sJt_{j} < 0$ this represents a physically unimportant
translation in $\theta$. The equation $\cd \Kbr/\cd\theta=0$ gives, for
$0\leq \theta < \pi$, $\theta=0$ and $\pi/2$. At $\theta=\pi/2$ the equation
$\cd\Kbr/\cd\Ie=0$ has a single root near $\Ie=\alpha_{j}$ and this fixed 
point is a centre. At $\theta=0$ there are generally three roots: 
a saddle near $\Ie=\alpha_{j}$,
but there are two others near $\Ie=\pm(\In-\Ism)$, because of the square-root
singularity in $A(\Ie)$. If the $(\theta,\Ie)$ phase plane is projected onto
a sphere with latitude $\psi$, so $\Ie=(\In-\Ism)\cos\psi$,  it is seen
that there are phase curves with centres close to, but not at, the
poles and which enclose the poles: in the Cartesian coordinate system 
$(\theta,\Ie)$  this produces the two extra fixed points. The physically
significant fixed points are near $\Ie=\alpha_{j}$ and these exist only
if $|\alpha_{j}| < \In-\Ism$, approximately.

By plotting the contours of $\Kbr$ for fixed $\Fm$ and $\Omega$ and with
$\Fs$ increasing so $\alpha_{j}$ increases from below $-(\In-\Ism)$ 
to above $\In-\Ism$ 
we see how the resonance develops and why, in certain circumstances, 
ionisation is enhanced if $\Fs\simeq \Fs^{(j)}$.

The following five figures show the contours of $\Kbr$, near the $j=1$ 
resonance for $\Ism=0.2$, $\omo=0.0528$ and $F_{\mu}=0.13$, corresponding to 
figure~\ref{f:2}. For these graphs we use
$\Kbr$ with $\gt_{1}$ and $\gb$ given by equations~\ref{eq:rh-02a} 
and~\ref{eq:rh-02} (to $O(F^{5})$) respectively; these give $\nu_{1}=0$ when 
$\Fs^{(1)}=0.0168$. 
In each figure $\Fs$ and $\Fmu$ are fixed so, according to the 
adiabatic ionisation criterion, there is a critical value of $\Ie$, 
above which orbits ionise, 
given by the solution of $\Fcrit(\Ie,\Ism)=\Fs+\Fmu$: for
the parameters used here the critical value of $\Ie$ changes from 
$\Ie=0.50$ ($\Fs=0.0152$) to $\Ie=0.42$  ($\Fs=0.0183$).
The maximum of $\Fcrit(\Ie,\Ism)$ is at $\Ie=1-|\Ism|$ so if
$\Fs+\Fmu < \Fcrit(1-|\Ism|,\Ism)$
there is no ionisation, even at a resonance. The upper solid horizontal line 
in each figure is at the value of $\Ie$ at which
$\Fcrit(\Ie,0.2)=\Fs+\Fm$, so orbits straying above this line will ionise.
The lower solid line is $\Ie=-0.4$  is taken, for illustrative purposes,
to be the initial state.
Note that in this case $\Fm-\Fs < 0.13$, so the other adiabatic boundary 
defined in equation~\ref{eq:st30} does not lead to ionisation.

\drlabel{f:10/11/12}
\begin{center}
\begin{fig}{100}{100}{mean-cont-f1a}{
}{\\$\alpha_{1}=-0.75,\,\Fs=0.0152$}{f:10}
\end{fig} \hspace{5pt}
\begin{fig}{100}{100}{mean-cont-f2a}{
}{\\$\alpha_{1}=-0.25,\,\Fs=0.0163$}{f:11}
\end{fig} \hspace{5pt}
\begin{fig}{100}{100}{mean-cont-f3a}{
}{\\$\alpha_{1}=0,\,\Fs=0.0168$}{f:12}
\end{fig}
\end{center}

\vspace{-30pt}

\drlabel{f:13/14}
\begin{center}
\begin{fig}{100}{100}{mean-cont-f4a}{
}{\\$\alpha_{1}=0.25,\,\Fs=0.0173$}{f:13}
\end{fig}\hspace{40pt}
\begin{fig}{100}{100}{mean-cont-f5a}{
}{\\$\alpha_{1}=0.75,\,\Fs=0.0183$}{f:14}
\end{fig}
\end{center}

\noindent
As $\Fs$ increases from $0.0152$ to $0.0183$, through $\Fs^{(1)}=0.0168$, the 
centre of the resonance island moves upwards. The physical effect of this is 
understood by considering a field suddenly switched on, with the initial
value of $\phie$ uniform in $(0,2\pi)$ and the initial value of $\Ie$, 
to be $-0.04$.

\begin{itemize}
\item Figure~\ref{f:10}: $\Fs=0.0152$ ($\alpha_{1}=-0.75$). The adiabatic 
condition shows that orbits for which $\Ie(t)>0.50$ ionise; at this field there
is no resonance island and no ionisation from the initial state. 

\item Figure~\ref{f:11}: $\Fs=0.0163$ ($\alpha_{1}=-0.25$). The 
resonance island exits; it intersects the initial state, but does not 
overlap the ionising region, so there is no ionisation. In practice the 
demarcation between ionising and non-ionising regions is less sharp because the
averaging approximations used to derive this simple picture replaces
unstable manifolds by separatrixes.

\item Figure~\ref{f:12}: $\Fs=\Fs^{(1)}=0.0168$  ($\alpha_{1}=0$). The
resonance island is in the centre of the phase space. The ionisation criterion 
is practically the same as in figure~\ref{f:10} so orbits with 
$\Ie(t)>0.50$ ionise. Now, however,
the resonance island can transport initial states to the 
ionising region, $\Ie> 0.46$. Note that not all orbits trapped in the
resonance island will ionise, but only those near the separatrix. We shall
see in section~\ref{sec:time-scale} how this affects the ionisation times.

\item Figure~\ref{f:13}: $\Fs=0.0173$  ($\alpha_{1}=0.25$). The centre of the
island is now at $\Ie=0.2$ and its separatrix just dips below the initial 
state, so few orbits ionise. In these circumstances it is shown in 
section~\ref{sec:envel} that the ionisation probability can be affected
significantly by the way the field is switched on.

\item Figure~\ref{f:14}: $\Fs=0.0183$  ($\alpha_{1}=0.75$). As for 
figure~\ref{f:10} the island no longer exists and there is no ionisation 
from the initial state. 
\end{itemize}

\noindent
This qualitative description of the ionisation process suggests that for 
a microcanonical distribution of substates the background ionisation 
increases as $\Fs$ increases across a resonance, because $\Ie^{c}$ decreases,
as shown in figure~\ref{f:w01}.

The centre of the $j$th resonance island is at approximately
\[
\Ie\simeq \alpha_{j}=\frac{4(3\Fs-j\omo)}{3(\Fs^{2}+\frac12\Fmu^{2})} 
\quad ({\rm in\;scaled\;units}),
\]
see equation~\ref{eq:st20a}, so it exists only for $\Fs$ in the interval
\drlabel{eq:st22}
\begin{equation}
\frac13 j\omo- \beta < \Fs < \frac13 j\omo+\beta,
\quad \beta=\frac{1}{36} j^{2}\omo^{2}+\frac18\Fmu^{2}.
\label{eq:st22}
\end{equation}
In this field range a high proportion of initial values of $\Ie(0)$ may 
lead to  ionisation: outside this interval the resonance does not exist. 

This qualitative description of the ionisation mechanism shows that for a 
system initially in a given $\Ism$-substate there are several conditions 
necessary for a resonance to enhance the ionisation probability.

\begin{enumerate}
\item[1)] The field amplitudes must be sufficiently large that there is 
ionisation for some value of $\Ie$, for the given $\Ism$.

\item[2)] The field amplitudes must not be so large that $\Pion=1$.

\item[3)] If $\Fcrit(\Ie^{c},\Ism)=\Fs+\Fm$, then the island width must 
exceed $\Ie^{c}-\Ie(0)$, otherwise the initial state cannot be transported
to an ionising state. There is, of course, a similar relation for the boundary
defined by $\Fcrit(-\Ie^{c},-\Ism)=\Fm - \Fs$ when $\Fm > \Fs$.
\end{enumerate}
The first two of these conditions define a region in the $(\Fs,\Fm)$-plane 
in which
the resonance may enhance ionisation. The boundaries of this region depend 
upon $\Ism$ and are the complement of the region defined by the two 
conditions $\Pion=0$ and $\Pion=1$ for all $\Ie$. Using the adiabatic 
assumption, $\Pion=1$ if $\Fs+\Fm>\max(\Fcrit)$ and $\Pion=0$ if
$\Fs+\Fm<\min(\Fcrit)$ and  $\Fm-\Fs < \min(\Fcrit)$, if $\Fs<\Fm$,
see equation~\ref{eq:st30}. In the
case $\Ism=0.1$ this region is shown by the shaded area in the following 
figure: outside this region the resonance can have no effect. If $|\Ism| > 0.1$
the equivalent region lies inside the area shown.


\drlabel{f:15a}
\begin{center}
\begin{fig}{252}{126}{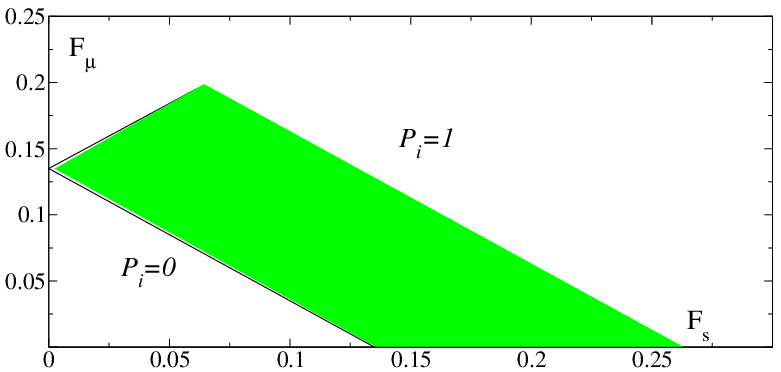}{}
{Diagram showing the regions where $\Pion=0$ and $\Pion=1$ for 
$\Ism = 0.1$. In the shaded area $0 < \Pion < 1$: only in this region are 
the resonances potentially visible.}{f:15a}
\end{fig}
\end{center}

Inside the shaded region a resonance affects the ionisation probability only if
condition~3 above is satisfied.
Below, but near, the upper boundary $\Pion \sim 1$, so resonance peaks are
barely noticeable, see for instance the $j=5$ resonance in figure~\ref{f:1}
at $\Fs\simeq 0.148$. The break down of adiabatic invariance 
broadens these boundaries but in a manner difficult to 
estimate, although the effect increases as $\omo$ increases; tunnelling 
also affects these boundaries.

There are three parameters of the resonant island that affect
$\Pion(\Fs)$. These are most easily estimated by setting $\Fs=\Fs^{(j)}$, so
$\alpha_{j}=0$, and $\Ism=0$ as well as using the lowest-order estimates 
of all variables: these values are used in the remainder of this section.

The first parameter is the island area, $\sA_{j}$, which determines how the
classical resonance affects the quantum dynamics. An estimate for this is,
\[
\frac{\sA_{j}}{2\pi}=\frac{2\In}{\pi} \sin^{-1} 
\sqrt{\frac{8\omo\Fmu |J'_{j}|}{3(\Fs^{2}+\frac12\Fmu^{2})+ 
4\omo\Fmu |J'_{j}|} },
\quad J'_{j}=J'_{j}(j\Fmu/\Fs), \quad \Fs=\Fs^{(j)}.
\]
For other values of $\Ism$ and $\alpha_{j}$ the form of the resonance
Hamiltonian  shows that $\sA_{j}$ is proportional to 
$\sqrt{|J'_{j}|}$.  
For the parameters of figure~\ref{f:1}, $\omo=0.098$, $\Fmu=0.1$ this gives
$\sA_{j}/2\pi=(0.80$, $0.05$, $0.35$, $0.25)\In$ for $j=1,\,2,\,3$ and 
$4$. In this case $\nze=43$ so 
the approximate number of states associated with these
islands are 34, 2, 15 and 11, respectively. Thus if $\nze$ is 
decreased, all other scaled variables remaining the same, we should expect,
in quantum dynamics, the $j=2$ resonance to become less prominent than the 
other resonances. 

The second parameter is the island width,
$\Delta\Ie$, that is the maximum distance between the two branches of the 
separatrix. It is difficult to derive a simple estimate of $\Delta\Ie$; here
we simply note that it is proprtional to $|J_{j}'|$, defined above.
A necessary condition for enhanced ionisation is that $\Delta\Ie$ 
is larger that $\Ie^{c}-\Ie(0)$; otherwise transport to ionising regions 
does not occur. Notice that this condition is independent of the principal
quantum number, $n$, unlike that discussed above.

The third important classical parameter is the period of the mean 
motion inside the 
island; as we shall see,  this determines how rapidly a resonance develops,
section~\ref{sec:time-scale}, and how it is affected by the field
envelope, section~\ref{sec:envel}. The frequency, $\omega_{j}$, of the 
motion inside the resonance island is approximated by
expanding the resonance Hamiltonian, equation~\ref{eq:st20a} about its centre
and near the island centre we obtain, in scaled units
\drlabel{eq:per01}
\begin{equation}
\omega^{2}_{j}=\frac38\left( \Fs^{2}+\frac12\Fmu^{2} \right)\omo\Fmu 
\left| J'_{j}\left(\frac{j\Fmu}{\Fs}\right)\right|,
\quad \Fs=\frac13 j\omo.
\label{eq:per01}
\end{equation}
This estimate gives the largest value of the frequency in 
the island; for motion nearer the separatrix $\omega_{j}$ is smaller. 

In this section we have shown how a dynamical resonance can enhance
the ionisation probability and have derived some approximate necessary 
conditions. The analysis uses the resonance Hamiltonian, $\Kbr$,
equation~\ref{eq:st20a}, derived using two stages of averaging. Moreover,
an important part of this Hamiltonian is the factor $\sJt_{j}$,
which, for small $\omo$, oscillates between its maximum and minimum values for
relatively small changes in $\Fm$. Hence, whilst $\Kbr$ provides a good
qualitative description, for any fixed $(\omo,\Fm)$ the details may be wrong;
for instance the field at which a resonance disappears is given inaccurately
by this approximation if $\omo$ is small.

\subsection{Resonance positions}
\drlabel{sec:res-ion-II}\label{sec:res-ion-II}
In this section we examine ionisation from a particular substate and 
compare theoretical predictions with exact numerical calculations.
We choose the low frequency $\omo=0.011414$,
$(\nze=21)$ (to minimise non-adiabatic effects), fix $\Ism=0.2$,
use the initial condition $\Ie=-0.4$ (so there is no average over 
substates) and put $\Fm=0.13$. 

Since $\Ie$ is an approximate constant of the motion and 
$\Fcrit(-0.4,0.2)=0.1984$, if the dynamics were adiabatic we should expect
complete ionisation when $\Fs$ exceeds $\Fcrit-\Fm=0.0684$ and no ionisation
for smaller static fields. At a resonance $\Ie(t)$ varies over part of its
accessible range and since $\min(\Fcrit)=0.1357$ we might
see the effect of the $j$th resonance if $\Fs^{(j)}\geq 0.0057$, provided
the size of the 
the resonance island is sufficiently large. This simple analysis suggests
that the resonances $2 \leq j \leq 18$ could be seen via ionisation: in 
practice, for reasons to be discussed later the $j=2,\cdots,6$, 11 and 14  
resonances are not observed.

In figure~\ref{f:n21a}  we show the classical ionisation
probabilities for the envelope 16-50-16 in which the $j=$7-10, 12, 13 
and 15-19 resonances are clearly visible, but the $j=6$, 11 and
14, marked by the arrows, are missing: other calculations show that the $j=5$ 
resonance is also missing and theory suggests that the $j=2$-4 resonance
islands are too narrow to affect the ionisation probability, that is
$\Delta\Ie < \Ie^{c}-\Ie(0)$, as discussed in the previous section.

\drlabel{f:n21a}
\begin{center}
\begin{fig}{240}{120}{n21a.eps}{
}{Ionisation curve for $\omo=0.011414$, $(\nze=21)$, $\Fm=0.13$, with
initial conditions $\Itw=0.2$, $\Ism=0.2$.}{f:n21a}
\end{fig}
\end{center}

\noindent
In table~\ref{t:10} are listed some parameters associated with the 
$j=7$-15 resonances. Here the resonance width, $\Delta\Fs^{(j)}$,
 is defined to be the difference
$\Fs^{+}-\Fs^{-}$ where $\Fs^{\pm}$ are respectively the smallest and 
largest values of $\Fs$ on either side of $\sF_{s}^{(j)}$ at which 
\mbox{$\Pion=0$}.
Notice that this width is generally less than the difference
$|\Fs^{(j)}-\sF_{s}^{(j)}|$. In these calculations 500
orbits were used. The values of $\sF_{s}^{(j)}$, the static field at
which $\Pion$ is largest, are computed using a grid $10^{-5}$ in
$\Fs$, and $\sF_{s}^{(j)}$ is taken to be at the maximum value of $\Pion$ on
this grid. The value of $\Fs^{(j)}$  is the root of 
$\gb(\Fs,\Fm)=j\omo/3$: below the double lines $\Fs^{(j)}+\Fm> F_{rc}$, the
approximate radius of convergence of the series for $\gb$, see
equation~\ref{eq:rh-01}.

\drlabel{t:10}
\begin{table}[htbp]
\caption[]{\small \label{t:10} Table showing the positions, heights and widths
of the resonances seen in figures~\ref{f:n21a}.
For completeness, the values of $\Fs^{(j)}$ are  0.00360, 0.00720, 0.01079, 
0.01438, 0.01795 and 0.2151, for $j=1$-6,  respectively.}

\vspace{2pt}

\begin{center}
\begin{tabular}{|c | ll | ll| }\hline
 & $\sF_{s}^{(j)}$ & $\Fs^{(j)}$ && \\ \hline
$j$ & Monte-Carlo & Pad\'{e} &  & Width  \\ 
 & estimate & approximate & $\Pion(\sF_{s}^{(j)})$ & $\Delta\Fs$ \\ \hline
7 & 0.02495 & 0.02506 & 0.38 & $10\times 10^{-5}$ \\
8 & 0.02841 & 0.02859& 0.27 & $7\times 10^{-5}$ \\
9 & 0.03194 & 0.03209 & 0.42 & $11\times 10^{-5}$ \\
10 & 0.03544 & 0.03556 & 0.68 & $31\times 10^{-5}$ \\
11 & ~~~-    & 0.03900 &  0   & $< 10^{-5}$ \\ \hline \hline
12 & 0.04235 & 0.04240 & 0.73 & $32\times 10^{-5}$ \\
13 & 0.04581 & 0.04576 & 0.73 & $46\times 10^{-5}$ \\
14 & ~~~-    & 0.04906 &  0    & $< 10^{-5}$ \\
15 & 0.05269  & 0.05229 & 0.94 & $77\times 10^{-5}$ \\ \hline
\end{tabular}
\end{center}
\end{table}

\noindent
The function $\gb(\Fs,\Fm)$ is known only via its series expansion, 
equation~\ref{eq:rh-02},
which has been computed to $O(F^{17})$, that is the first nine terms.
From the discussion after equation~\ref{eq:rh-01}, since $\max(F)=\Fs+\Fm$, we 
expect any theory based on the series representation of $\gb$ to be 
valid for those resonances satisfying $\Fs^{(j)} +\Fm < 0.18$ ($\Ism=0$) 
and $0.22$~($\Ism=1$).

It transpires that if $F$ is near the upper boundary the
values of $\Fs^{(j)}$ are sensitive to the 
number of terms in the series for $\gb$ and extrapolation is necessary
to estimate converged values. Here we consider two methods 
of extrapolating and give reasons which suggest that the  Pad\'{e} 
approximant is more reliable. All the following results 
are obtained by substituting 
$\Fm=0.13$ and $\Ism=0.2$ into the series for $\gb$ and then manipulating 
the resultant power series in $\Fs$: for completeness we give this series:
\drlabel{eq:pos05}
\begin{eqnarray}
\frac{\gb(\Fs)}{\Fs} &=&1.057 + 0.08759x + 0.1079x^{2}+
0.1073 x^{3}+ 0.05903x^{4}+ 0.01588x^{5} \nonumber \\
&& \hspace{2pt} + 0.001861 x^{6}+ 7.700\times 10^{-5}x^{7}+ 
6.509\times 10^{-7}x^{8}, \; x=(10\Fs)^{2}.
\label{eq:pos05}
\end{eqnarray}
For the $j=7$ resonance the lowest order approximation
gives $\Fs^{(7)}\simeq 7\omo/3=0.02663$, which is about 7\% too large. Eight 
other estimates can be obtained by truncating the series for $\gb(\Fs)/\Fs$ 
at $\Fs^{2k}$,
$\range{k}{1}{2}{8}$: these are 0.02566, 0.02536, 0.02522, 0.02515, 0.02511, 
0.02509, 0.02507 and 0.2506. This sequence appears to be converging, 
but has not 
reached its limit (to the four significant figures quoted). Suppose we 
have $M$ estimates $\sF_{p}$ for $\Fs^{(j)}$, using $\range{p}{1}{2}{M}$
terms of the series~\ref{eq:pos05}, then using the method of Richardson
we assume that $\sF_{p}=\Fs^{(j)}+\sum_{r=1}^{M-1}A_{r}p^{-r}$. 
These $M$ equations may be solved for the unknown $\Fs^{(j)}$
to give the estimate $\Fs^{(7)}\simeq 0.02505$, which
differs from $\sF_{s}^{(j)}(=0.02495)$ by $10^{-4}$. Despite this 
relatively small inaccuracy we note that a Monte-Carlo calculation 
(with 500 orbits)
gives $\Pion(0.02505)=0$, with $\Pion(\Fs)\neq 0$ in the interval
$0.02495\pm 5\times 10^{-5}$, which does not overlap with our estimate
of $\Fs^{(7)}$.

Another approach is to form a Pad\'{e} approximant of $\gb(\Fs)/\Fs$
using the expression~\ref{eq:pos05}, treated as an eighth degree polynomial
in $\Fs^{2}$. The coefficients, and hence the positions of the poles, of 
these approximants depend upon $\Ism$ and $\Fm$: we find that the position of 
the pole nearest the origin is relatively insensitive to $\Ism$, but changes
significantly with $\Fm$. Therefore for $\Fm < 0.09$ we use a $[2/2]$ 
approximant (in $\Fs^{2}$) and for $\Fm\geq 0.09$ we use a $[3/3]$ approximant.

For the case considered here, $\Fm=0.13$, $\Ism=0.2$, the relevant Pad\'{e} is
\drlabel{eq:pos06}
\begin{equation}
\frac{\gb}{\Fs}=\frac{1.057-9.542x+64.89x^{2}-156.4x^{3}}
{1-9.858x+59.36x^{2}-198.0x^{3}},\quad x=(10\Fs)^{2}.
\label{eq:pos06}
\end{equation}
With this approximation for $\gb(\Fs)$ the equation $3\gb=j\omo$ gives
the values of $\Fs^{(j)}$ quoted in table~\ref{t:10}.

For the $j=15$ resonance the zero-order approximation to $\Fs^{(15)}$ is
$0.0571$, which is about 8\% too large. The eight other estimates, obtained 
using the truncated series for $\gb(\Fs)$, equation~\ref{eq:pos05}, are
0.05479, 0.05395, 0.05347, 0.05313, 0.05287, 0.05265, 0.05246 and 0.05229. 
Richardson's extrapolation gives $\Fs^{(15)}=0.04858$, 
which is 8\% smaller than $\sF_{s}^{(j)}(=0.05265)$, and no improvement on
the zero-order approximation. The $[3/3]$ Pad\'{e} approximant, 
equation~\ref{eq:pos06}, gives $\Fs^{(15)}=0.05229$, approximately $0.7\%$
smaller than $\sF_{s}^{(j)}$. In this case the Pad\'{e} approximant seems 
to provide a 
more reliable method of extrapolating the truncated series for $\gb(\Fs)$.
We note that this resonance field is on the edge of the validity
of the series expansion, so any estimate of $\Fs^{(15)}$ based on the 
series may not be accurate; in these circumstances, however, the Pad\'{e}
approximant is more likely to provide an accurate estimate of the exact
function. In all future estimates of $\Fs^{(j)}$ we therefore use the
Pad\'{e} approximant.

We now turn our attention to the resonances missing from
figure~\ref{f:n21a}. For this analysis we use the resonance Hamiltonian
used to plot the contours in figures~\ref{f:10}-\ref{f:14} which, for the
reasons discussed at the end of the previous section,
provides only a qualitative description.

Using the simple approximations for $\gb$ and $\gt_{1}$, 
equations~\ref{eq:rh-02} and~\ref{eq:rh-02a}, we find than the width
of the resonance island is proportional to
\drlabel{eq:pos07}
\begin{equation}
A(j)=J_{j}'\left( \frac{3\Fm}{\omo}\gt_{1}(\Fs^{(j)},\Fm)\right)
\quad {\rm where} \quad 
\gb(\Fs^{(j)},\Fm)=\frac13 j\omo.
\label{eq:pos07}
\end{equation}
An overview of the widths of the $j=2$-15 resonances is given by the
graph of this function with $j$ taking all real values. This graph is
shown  next, with integer values of $j$ being marked by the circles.

\begin{center}
\drlabel{f:res-diss}
\begin{drfigs}
\drsetfig{280pt}{res-diss-fig02.eps}{Graph of the function $A(j)$ for
$\Fm=0.13$, $\omo=0.011414$ and $\Ism=0.2$}{f:res-diss}
\end{drfigs}
\end{center}

\noindent
We relate this graph to the ionisation curve in figure~\ref{f:n21a} by
recalling that a dynamical resonance affects the ionisation probability only
if it can transport an orbit to a region 
$\Ie > \Ie^{c}$, where $\Ie^{c}$ is defined by 
$\Fcrit(\Ie^{c},\Ism)=\Fm+\Fs^{(j)}$, so the resonance island width must
exceed the difference $\Ie^{c}-\Ie(0)$.

The values of $\Ie^{c}\left(\Fs^{(j)}\right)$ decrease with 
increasing $j$; for $j=2$,
$\Ie^{c}=0.75$ and for $j=5$ $\Ie^{c}=0.43$, but for these cases the maximum
possible size of the resonance island is smaller than $\Ie^{c}-\Ie(0)$.
For $j=6$, 11 and 14  the resonance island is seen from figure~\ref{f:res-diss}
to be very small so these resonances do not affect the ionisation probability.

For $j=7$-10 the simple Hamiltonian with contours shown in 
figures~\ref{f:10}-\ref{f:14} suggests that the resonance island is slightly 
too small for enhanced ionisation. But non-adiabatic effects, the 
approximations used --- recall that two averaging approximations have been used
to derive this Hamiltonian --- and the field envelope will broaden these
boundaries. The $j=11$ and 14 resonance islands are predicted to be too 
small to promote ionisation, whereas this simple approximation predicts
enhanced ionisation for all other $j$.

\subsection{Resonance disappearance}
The $j$th resonance has no effect on the dynamics at those values
of $\Fm$ where $\sJ_{j}=0$, see equation~\ref{eq:st20a}.
With $\gb(\Fs,\Fm)$ approximated by a Pad\'{e} approximant, as in 
equation~\ref{eq:pos05}, the equation $\gb=j\omo/3$ provides an expression for 
$\Fs^{(j)}(\Fm)$ and then equation~\ref{eq:st18b} can be used to 
obtain the numerical value of $\Fm^{(j,k)}$, associated with 
$j'_{j\hsp{0.5}k}$.
In the following table we give values of $\Fs^{(j)}$ and $\Fm^{(j,k)}$,
nearest 0.13, at the $j=5$-15
resonances for $\omo=0.011414$. In this example 
the difference between $\Fm^{(j,k+1)}$ and  $\Fm^{(j,k)}$ is about~0.01.

\drlabel{t:16}
\begin{table}[htbp]
\caption[]{\small \label{t:16} Table showing the value of $\Fs^{(j)}$ and
$\Fm^{(j,k)}$ nearest $0.13$, at which each resonance shown in 
figure~\ref{f:n21a}, where $\Fm=0.13$, disappears.}

\vspace{4pt}

\begin{center}
\begin{tabular}{|l c c c c c | }\hline
\hsp{8} $j$ & 6 & 7 & 8 & 9 & 10 \\ \hline
$\Fs^{(j)}$ & 0.0215 &0.0249 & 0.0287 & 0.0319 & 0.0358  \\
$\Fm^{(j,k)}$ & 0.130 & 0.135 & 0.128 & 0.133 & 0.126  \\ \hline \hline
\hsp{10} $j$ & 11 & 12 & 13 & 14 & 15 \\ \hline
$\Fs^{(j)}$ & 0.0390 & 0.0420 & 0.0460 & 0.0489 &0.0515 \\ 
$\Fm^{(j,k)}$ & 0.130 & 0.135 & 0.127 & 0.131 & 0.136 \\ \hline
\end{tabular}
\end{center}
\end{table}

\noindent
Observe that for $j=6$, 11 and 14, $\Fm^{(j,k)}\simeq 0.13$ and that for
$j=8$ it is close to $0.13$ and at this resonance $\max(\Pion)$ is relatively
small.

In the experimental results reported by Galvez {\em et al} (2000) it was shown
that resonances disappear at certain field values 
$(\sF_{s}^{(j)},\Fm^{(j,k)})$, given approximately by 
equation~\ref{eq:st20b}. Since then two sets of more accurate measurements 
have been made. First, disappearances in the 8.105 Ghz cavity, with scaled 
frequencies mostly in the range $0.0731 \leq \omo \leq 0.136$, are reported 
by Galvez {\em et al} (2004). Second, Schlultz (2003) has reported
results for a cavity with frequency 3.5539 GHz, and scaled frequencies in the
range $0.035 < \omo < 0.16$, see also Schultz {\em et al} (2004). 
In those papers experimental values of 
$(\sF_{s}^{(j)},\Fm^{(j,k)})$ are compared with classical Monte 
Carlo estimates: here we compare some of these with theoretical values.

With the Monte-Carlo method it is not feasible to compute the exact
values of $(\sF_{s}^{(j)},\Fm^{(j,k)})$, so we determine
an interval for each of $\Fs$ and $\Fm$ in which the resonance cannot be
distinguished from the statistical fluctuations. The mid point of this 
rectangle is taken to be the point of disappearance. 

For the 29 cases considered by Schultz (2003) the relative difference between 
the experimental and
computed values of $\sF_{s}^{(j)}$ is less than 1\% for 22 cases and between
1\% and 2\% in 5 cases: for $\Fm^{(j,k)}$ the corresponding percentages 
are 11 and 16 respectively.
For a comparison with theory we compute 
$(\Fs^{(j)},\Fm^{(j,k)})$ for $\Ism$ between 0
and 1 and compare the average with the experimental values of 
$(\sF_{s}^{(j)},\Fm^{(j,k)})$. Of the 32 cases considered the 
relative difference between $\Fs^{(j)}$ and $\sF_{s}^{(j)}$  is less than 
1\% for 18 cases and between 1\% and 2\% in 10 cases: for $\Fm^{(j,k)}$ the 
corresponding percentages are 24 and 6 respectively.

Some typical comparisons between the Monte Carlo calculations and the 
theoretical values, for various scaled frequencies, are shown in 
table~\ref{t:12}.

\drlabel{t:12}
\begin{table}[htbp]
\begin{center}
\caption[]{\small \label{t:12} Comaprison of the theoretical values of
$(\Fs^{j)}, \Fm^{(k,j)})$ and the Monte Carlo estimates of 
$(\sF_{s}^{(j)},\Fm^{(j,k)})$ for various scaled frequencies and 
values of $j$. In the extreme right column is
$\sF_{s}^{(j)}+\Fm^{(k,j)}$ because the discussion after 
equation~\ref{eq:rh-01} 
suggests that this theory will be unreliable if this value exceeds 0.19.}

\vspace{4pt}

\begin{tabular}{| l | l |rr | ll | c |} \hline
$\omo$ & $j$ & $\sF_{s}^{(j)}$  & $\Fs^{(j)}$ & $\Fm^{(k,j)}$ 
& $\Fm^{(k,j)}$ & $\sF_{s}^{(j)}+\Fm^{(k,j)}$  \\ \hline
       &     &  MC~~    & Theory      & MC & Theory & \\ \hline
0.0367 & 1 & 0.0115 & 0.0113 & 0.139 & 0.140 & 0.14 \\
0.0789 & 1 & 0.0242 & 0.0244 & 0.134 & 0.137 & 0.16 \\ \hline
0.0980 & 2 & 0.0623 & 0.0626 & 0.0949 & 0.0962 & 0.16 \\
0.136  & 2 & 0.0832 & 0.0778 & 0.127 & 0.131 & 0.21 \\ \hline
0.0789 & 3 & 0.0740 & 0.0739 & 0.104 & 0.105 & 0.18 \\ \hline
0.0731 & 4 & 0.0895 & 0.0852 & 0.120 & 0.121 & 0.21 \\ 
0.0789 & 4 & 0.0957 & 0.0868 & 0.128 & 0.130 & 0.22 \\ \hline \hline
\end{tabular}
\end{center}
\end{table}

\subsection{Variation of resonance position with $\Ism$}
Now consider the variation of $\Fs^{(j)}$ with $\Ism$. From 
the series for $\gb(F)$, equation~\ref{eq:rh-02}, we see that 
$\Fs^{(j)}$ decreases as $\Ism$ increases, but that the difference 
between the largest and smallest values is approximately
\(
11j\omo(9\Fm^{2}+(j\omo)^2)/432
\):
for the data in table~\ref{t:10} this gives $4.4j\times 10^{-5}$
approximately. The results obtained using the Pad\'{e} approximant for $\gb$
are given in table~\ref{t:11}. 
Columns~2 to~4 show the mean $\mean{\Fs^{(j)}}$, averaged 
over $\Ism$, and the minimum and maximum values of $\Fs^{(j)}$, 
for $\Fm=0.13$ and $\omo=0.0114$.

\drlabel{t:11}
\begin{table}[htbp]
\begin{center}
\caption[]{\small \label{t:11} Table of the theoretical resonance
positions for the parameters defined in table~\ref{t:10}, 
$\Fm=0.13$ and $\omo=0.0114$. In column~5 is 
the difference between the largest and smallest values of $\Fs^{(j)}$. }

\vspace{4pt}

\begin{tabular}{|c|c|c|c|c|} \hline
$j$ & $\mean{\Fs^{(j)}}$ & $\min(\Fs^{(j)})$ & $\max(\Fs^{(j)})$ & 
spread \\
 & & $\Ism=1$ &$\Ism=0$ & $\delta F^{(j)}$ \\ \hline
5  & 0.01788 & 0.01771 & 0.01796 & $25\times 10^{-5}$ \\
6  & 0.02143 & 0.02123 & 0.02153 & $29\times 10^{-5}$ \\
7  & 0.02496 & 0.02475 & 0.02507 & $33\times 10^{-5}$ \\
8  & 0.02848 & 0.02825 & 0.02860 & $35\times 10^{-5}$ \\
9  & 0.03198 & 0.03173 & 0.03210 & $37\times 10^{-5}$ \\ 
10 & 0.03545 & 0.03520 & 0.03558 & $38\times 10^{-5}$ \\
11 & 0.03890 & 0.03864 & 0.03902 & $39\times 10^{-5}$ \\
12 & 0.04231 & 0.04207 & 0.04241 & $34\times 10^{-5}$ \\
13 & 0.04568 & 0.04547 & 0.04576 & $29\times 10^{-5}$ \\
14 & 0.04901 & 0.04884 & 0.04906 & $22\times 10^{-5}$ \\
15 & 0.05228 & 0.05218 & 0.05229 & $13\times 10^{-5}$ \\ \hline \hline
\end{tabular}
\end{center}
\end{table}

\noindent
This data shows that, for this low frequency, the above simple estimate of 
the spread is reasonable for $j \leq 10$, and that it is
comparable with the difference between the $\sF_{s}^{(j)}$ and $\Fs^{(j)}$, 
given in columns 2 and 3 of table~\ref{t:10} and to the resonance width, 
column~5. Note also that as $j$ increases, and $\Fs^{(j)}+\Fm$ 
tends towards the radius of convergence of the series for $\gb$, the 
difference decreases: it is not known if this effect is real or due
to the approximations used.

Although the position of the dynamical resonance is very weakly dependent
upon $\Ism$ its effect on $\Pion(\Fs)$ can depend strongly upon 
$\Ism$, because $\Ie^{c}$ --- that is the solution of 
$\Fcrit(\Ie^{c},\Ism)=\Fm+\Fs^{(j)}$ --- depends upon $\Ism$: for instance 
with $j=10$, $\Ie^{c}$ varies from $0.04$ ($\Ism=0$) to $0.3$ ($\Ism=1$).

\subsection{Resonance widths}
The shapes of the resonances seen in the experimental data and the 
classical simulations are complicated. In particular the resonances are
not normally symmetrical about $\sF_{s}^{(j)}$, with  details depending upon 
$\Fm$ and $j$. Here we show show the classical 
widths can be defined and computed and discuss some of the reason for the 
shapes observed. A detailed comparison between the 
classical simulations and the experimental data is provided in the companion
paper, Galvez {\em et al} (2004), and there it is shown that the experimental 
results display the same complexities.

In order to estimate the widths it is necessary to
isolate the resonances from the background. This is achieved by noting 
that on either side of the resonance $\Pion(\Fs)$ increases approximately
linearly, figure~\ref{f:w01} below. The background may be eliminated 
by subtracting these straight line segments from $\Pion(\Fs)$ to give an
adjusted probability that is approximately zero on both sides of the 
resonance, as shown on the right of figure~\ref{f:w01}. 
The only complication with this 
procedure is that the straight line segments have different gradients, so we 
form a new fit to the background, $\Pfit=m(\Fs)\Fs + c(\Fs)$, where the
gradient, $m(\Fs)$, and constant, $c(\Fs)$, change smoothly between the 
values either side of the resonance, 
$(m_{1},c_{1})$ and $(m_{2},c_{2})$ respectively.
If the straight line segments are on the intervals 
$F_{1}\leq \Fs\leq F_{2}$ and $F_{3}\leq \Fs\leq F_{4}$, with $F_{2}<F_{3}$,
(chosen by eye) we set
\[
m(F)=\frac{m_{1}-m_{2}}{2} + \frac{m_{1}+m_{2}}{2}\tanh(\alpha F+\beta),
\quad \alpha=\frac{4}{F_{3}-F_{2}}, \quad 
\beta=-\frac{2(F_{3}+F_{2})}{F_{3}-F_{2}},
\]
with a similar fit for $c(F)$.

In the following
two figures we show how this process works for the $j=2$ resonance with  
$\nze=47$ ($\omo=0.1278$) and $\Fmu=0.1$, which is a fairly typical example:
in these calculations a microcanonical distribution of initial states
with 1296 orbits, for each value of $\Fs$, is used and the envelope 
is 16-113-16.


\drlabel{f:w01}
%
\begin{center}
\begin{fig}{360}{120}{width-fig.eps}{}
{A graph of the $j=2$ resonance for $\omo=0.1278$ $(\nze=47)$ with $\Fmu=0.1$.
On the left is the ionisation curve with the straight 
line fits, as described in the text. On the right is the difference, having 
subtracted the background.}{f:w01}
\end{fig}
\end{center}

\noindent
The left hand panel of figure~\ref{f:w01} shows that  either side of a 
resonance $\Pion(\Fs)$ is approximately linear, but with different gradients.
The difference between these gradients changes with $\Fmu$ and, of course, 
is zero when the resonance disappears. The right hand panel shows
the graph of $\Pion(\Fs)-\Pfit(\Fs)$ which highlights the resonance 
shape; this is clearly asymmetrical about the maximum. The graph shown
is typical though the degree of asymmetry changes with $j$ and~$\Fmu$. The 
position of the resonance, the two gradients $m_{1}$ and 
$m_{2}$ and the width of the adjusted ionisation probability all provide  
tests for any theory. Comparisons of these parameters obtained from classical
calculations and experiment are given in Galvez {\em et al} (2004).

In this particular example $\sF_{s}^{(2)}=0.0802$: the 
width at half-height is about $\Delta\Fs=0.0018$, with the base 
being nearly 4 times wider, 0.007. The  calculated position of this dynamical
resonance varies from $\Fs^{(2)}=0.0797$ $(\Ism=1)$ to $0.0800$ $(\Ism=0)$.
The spread in $\Fs^{(2)}$ due to the variations in $\Ism$ is therefore
about $0.0003$, about $\frac16$ of the $\frac12$-height width seen in
figure~\ref{f:w01}: this difference is fairly typical.




We now consider some of the factors determining the resonance shapes, and
show how these are partly determined  by a combination of substate averaging 
and non-adiabatic effects. Consider
ionisation from a given substate:  for illustrative purposes choose
$\nze=47$ ($\omo=0.1278$), $\Ism=0$, $\Fm=0.1$ and use a 4-50-4 envelope
with initial conditions $\Ie(0)=-0.9$, $-0.8$, $-0.6$ and $-0.4$.
The values of  $\Fcrit -\Fm$ are depicted by the arrows in each of the 
four graphs and  adiabatic invariance suggests 
that $\Pion=1$ to the right of these arrows,

\begin{center}
\drlabel{f:w03}
\begin{drfigs}
\drsetfig{324pt}{fig-n47-subst01.eps}
{Ionisation probabilities for $\omo=0.1278$, $\Fm=0.1$, $\Ism=0$ and 
$\Ie(0)=-0.9$ and $-0.8$.}{f:w03}
\end{drfigs}
\end{center}

\noindent
When $\Ie(0)=-0.9$ the $j=1$-5 resonances are clearly present; the positions 
and full-widths, as
defined in section~\ref{sec:res-ion-II}, are given in table~\ref{t:15}. 
For $\Ie=-0.8$ only the first three resonances are visible because 
$\Fcrit=0.256$. In the next two figures
we see the $j=2$ resonance disappearing as $\Ie(0)$ increases.

\begin{center}
\drlabel{f:w04}
\begin{drfigs}
\drsetfig{324pt}{fig-n47-subst02.eps}
{Ionisation probabilities for $\omo=0.1278$, $\Fm=0.1$, $\Ism=0$ and 
$\Ie(0)=-0.6$ and $-0.4$.}{f:w04}
\end{drfigs}
\end{center}

\drlabel{t:15}
\begin{table}[htbp]
\caption[]{\label{t:15}\small Values of $\Fs^{(j)}$, $\sF_{s}^{(j)}$ 
and the full resonance width, for the resonances shown in 
figures~\ref{f:w03} and~\ref{f:w04}, the latter two items being computed as in 
table~\ref{t:10}, with the $\Fs$-grid being $10^{-4}$. The radius of 
convergence of the series~\ref{eq:rh-02}, for this problem, is about 0.18.}
\begin{center}
\begin{tabular}{| r | cccc | ccc | } \hline
$j$ \hsp{10}& 1 & 2 & 3 & 4 & 1 & 2 & 3 \\ \hline
 & \multicolumn{4}{c}{$\Ie=-0.9$} & \multicolumn{3}{c}{$\Ie=-0.8$} \\ \hline
$\Fs^{(j)}$ & 0.0413 & 0.0801 & 0.109 & 0.122 & 0.0413 & 0.0801 & 
0.109 \\ 
$\sF_{s}^{(j)}$  & 0.0407 & 0.0804 & 0.119 & 0.154 &0.0407 & 0.0805 & 
0.1185 \\
full-width & 0.0014 & 0.0018 & 0.0028 & &0.0017 & 0.0027 & 0.009 \\ \hline
\multicolumn{7}{c}{} \\ \hline
 & \multicolumn{4}{c}{$\Ie=-0.6$} & \multicolumn{3}{c}{$\Ie=-0.4$} \\ \hline
$\sF_{s}^{(j)}$ & 0.0408 & 0.0804 & --  & --& 0.0408 & 0.0788 & -- \\
full-width & 0.0026 & -- &  & & 0.0035 & -- &  \\ \hline
\end{tabular}
\end{center}
\end{table}

\noindent
In this example adiabatic invariance predicts that the 
$j=4$, 3, 2 and 1 resonances disappear when 
$\Ie=-0.80$, $-0.60$, $-0.20$ and $0.63$, respectively:
figures~\ref{f:w03} and~\ref{f:w04} show these
predictions to be approximately true. This data also shows that as $\Ie$ 
increases and the adiabatic boundary encroaches upon each resonance it
broadens, acquires an asymmetry and eventually disappears.
In other cases, when $\Fm>\Fs$, an ionisation boundary can also 
encroach from the left, figure~\ref{f:8}, and this will also
change the resonance shape.

This example, which is typical, suggests that the width and shape of the 
microcanonical averaged resonances is caused mainly by the effect of the 
separatrix between bound and free motion, which distorts
nearby resonances, rather than the variation in the resonance position
with $\Ism$. It is therefore difficult to provide theoretical estimates of 
the resonance width and shape.

\section{Resonance time-scales}
\drlabel{sec:time-scale}\label{sec:time-scale}
The classical adiabatic ionisation mechanism, described in 
section~\ref{sec:ion}, suggests that, in the absence of a resonance, 
ionisation 
occurs when $F=\lambda(t)(\Fs+\Fm)$ reaches a critical value defined 
by the condition $F=\Fcrit(\Ie,\Ism)$: at this
time ionisation from a particular orbit occurs within a Kepler period. This
behaviour has been checked numerically when $\Fs=0$ (Rath, 1990) and
the results of the present calculations, where $\Fs >0$, show the same 
behaviour. With increasing scaled frequency the dynamics becomes less
adiabatic although this behaviour persists, albeit with the boundaries
becoming blurred; this behaviour is seen clearly in figure~2 of 
Richards~(1996a).

The classical ionisation mechanism at a resonance is different because 
here for an 
orbit to be  ionised it must first be transported into a region of large
$\Ie$, that is smaller $\Fcrit$, by motion round the resonance island. Hence 
the rate of 
ionisation will depend upon the period of this motion; at
the island centre this is given approximately by equation~\ref{eq:per01},
which shows the period to be $O(F^{-3/2}\omo^{-1/2})$. Thus we should expect
the time dependence of the ionisation probability on and off 
resonance to be quite different.

These predictions can be checked by computing the time at which $\Pion(\Fs,t)$
reaches a given proportion of its final value.
In the following figure we show two graphs which allow comparison of this
ionisation time with the ionisation probability. The upper graph is the
ionisation probability, $\Pion(\Fs)$, for $\omo=0.0528$ $(\nze=35)$, $\Fm=0.13$
starting in the initial state $(\Ie,\Ism)=(-0.4,0.2)$, using a 16-50-16
envelope and 1600 orbits, which is the same as in figure~\ref{f:2}:
the lower graph is the time, 
$T_{h}$, at which $\Pion(\Fs,t)$ reaches half its final value: with
$T_{h}$ is measured in units of the field period: the
horizontal line is at $T=16T_{f}$, the time when the field amplitude
reaches its maximum.

\vspace{15pt}

\begin{center}
\drlabel{f:time}
\begin{drfigs}
\drsetfig{300pt}{n35-fm0p13-time-24jul03.eps}{
Ionisation probabilities, upper graph, and ionisation times, lower
graph, for the parameters defined in the text. Here 
$\omo=0.0528$~and~$\Fm=0.13$.}{f:time}
\end{drfigs}
\end{center}

\noindent
There are several features of this comparison worthy of note.
\newcounter{note}
\begin{list}
{(\roman{note})}{\usecounter{note}
\setlength{\labelsep}{3pt}%
\setlength{\labelwidth}{10pt}%
\setlength{\listparindent}{10pt}}%
\item For $\Fs > 0.06$ ionisation occurs close to end of the switch-on
time, $T=16T_{f}$. Since $\Fcrit(-0.4,0.2)=0.198$ the adiabatic condition 
suggests, that away from resonances and these initial conditions, bound 
states exist only for $\Fs < 0.07$.

\item At the $j=1$, 2 and 3 resonances, ionisation occurs some time after
the field has reached its maximum amplitude, with the longest delay
occurring at the edges of the resonance and the shortest near the maximum
in $\Pion$. This is consistent with the description given in 
section~\ref{sec:res-ion-I}, where it is shown that close to the resonance 
edge transport is near the separatrix where motion is slowest. The 
formula~\ref{eq:per01} for $\omega_{j}$, gives, for these parameters,
the $1/2$-period near the resonance island centre of about $14T_{f}$,
which is consistent with the lower graph of figure~\ref{f:time}.

\item The local maximum in $T_{h}$ at $\Fs\simeq \sF_{s}^{(2)}$ cannot,
at present, be explained.

\item The ionisation time near the local maximum in $\Pion$ at $\Fs=0.0428$ 
has the same shape as those near the $j=1$ and $2$ resonances, but ionisation
clearly takes longer, suggesting that this structure is due to a 
higher-order resonance. Linear interpolation between the $j=1$-3 resonances 
suggests
this could be the $j=2\frac23$ resonance. A similar calculation suggests that
local maximum in $T_{h}$ at $\Fs\simeq 0.0537$ could be the $j=3\frac13$ 
resonance.

It is not easy to see what produces these non-integer resonances. Second-order
perturbation theory applied to the mean motion Hamiltonian~\ref{eq:st15}
does not appear to give $1/3$ resonances; this suggests that higher 
harmonics of $\phie$  are required and these occur, at this level of 
averaging, only if higher order terms of $\thetaon(\psi,\chi)$ and
$\thetatw(\psi,\chi)$, equations~\ref{eq:2-10a} and~\ref{eq:app-ang04},
are included.

\item For $\Fs \geq 0.05$ the boundary at $\Fcrit-\Fm=0.068$ is begining to 
affect the dynamics and seems to be interfering with the $j=3$ resonance.
\end{list}

\section{Envelope effects}
\drlabel{sec:envel}\label{sec:envel}
In the previous section it was shown how resonance islands affect
ionisation times. Here we examine the effect of the envelope 
switch-on time, $T_{a}=2\pi N_{a}/\Omega$, on a particular
resonance. At this point it is useful to recall that the dynamical resonances
discussed here are unusual because each exist only for a narrow range
of $\Fs$ and within this interval the resonance island moves from the 
lower to the upper
edge of phase space, see figures~\ref{f:10}-\ref{f:14}. Moreover, the motion 
inside a resonance island is very slow, equation~\ref{eq:per01}. 
If $\Fs\sim\Fs^{(j)}$, for some $j$, then for most of the 
switch-on period the resonance island does not exist. But for some time 
close to $T_{a}$ the island develops at the bottom edge of phase space
and as $t\to T_{a}$ it moves up through phase space and through the 
initial phase line. As this happens the line is distorted, with the amount of
distortion depending upon the relative values of $d\lambda/dt$ and 
the frequency of the motion in the island. For short switch times
the initial phase line $\Ie(0)$ evolves into a nearby line at 
$t=T_{a}$, as shown in the left panel of figure~\ref{f:separ} below. For 
relatively long switch times there may be enough time for 
the initial phase line to develop an incipient homoclinic tangle and become
quite complicated, as seen in the right panel of figure~\ref{f:separ}. The
examples considered next show how changes in the switch-time, $T_{a}$, can 
dramatically affect the ionisation probability.

The demonstration of this effect is in two parts. First we show some exact
numerical results illustrating how the ionisation probability  changes 
with $T_{a}$. In this example we chose the low frequency,
$\omo=0.011414$, ($\nze=21$), $\Ism=0.2$, $\Fm=0.13$ and examine 
$\Pion(\Fs)$ in the vicinity of the $j=7$ resonance for the envelope 
$N_{a}$-50-$N_{a}$, for 
$N_{a}=1$--40. From table~\ref{t:10} we see that when $N_{a}=16$, 
$\Pion(\Fs)$ has its maximum at $\Fs=0.024950$ and that the resonance width
is $\Delta\Fs\simeq 8\times 10^{-5}$.

In the following two figures we show how $\Pion(\Fs)$ changes with $N_{a}$: for
these calculations we used 900 orbits and a grid $\delta\Fs=2\times 10^{-6}$,
so there are 50 data points for each unit of $f=(\Fs-0.024925)10^{4}$.

The first figure shows ionisation probabilities for $N_{a}=1$-17.
The variation of $\Pion$ with $N_{a}$ shows a surprising amount of 
variation; in particular we note that for $N_{a}=15$ the ionisation
probability is {\em zero} across the resonance. 

Also observe that $\sF_{s}^{(j)}$ changes by 
$\Delta\sF\sim 5.2\times 10^{-5}$ for
$N_{a}$ in this range, and that this is comparable to the resonance
width. This explains why an unambiguous relation between 
$\Fs^{(j)}$ and $\sF_{s}^{(j)}$ does not exist.


\drlabel{f:18a}
\begin{center}
\begin{drfigs}
\drsetfig{324pt}{switch-var01.eps}{Some graphs of the 
ionisation probability in terms of the scaled field 
\mbox{$f=(\Fs-0.024925)10^{4}$}
across the $j=7$ resonance for $\omo=0.011414$ and 
switch times for $1 \leq N_{a} \leq 17$.}{f:18a}
\end{drfigs}
\end{center}

\noindent
In the next two figures are shown the ionisation probabilities for
$23 \leq N_{a} \leq 40$. With these longer switch times more structure is
seen. For instance, with $N_{a}=31$ and 36, $\Pion(\Fs)$ has two local 
maxima and
for both $N_{a}=31$ and 37 the probability has a long, low plateau after the 
maximum. A qualitative explanation of these feature is given next.

\drlabel{f18b}
\begin{center}
\begin{drfigs}
\drsetfig{324pt}{switch-var02.eps}{Some graphs of the 
ionisation probability in terms of the scaled field 
\mbox{$f=(\Fs-0.024925)10^{4}$}
across the $j=7$ resonance for $\omo=0.011414$ and 
switch times for $23 \leq N_{a} \leq 40$.}{f:18b}
\end{drfigs}
\end{center}

\noindent
The behaviour depicted in figures~\ref{f:18a} and~\ref{f:18b} can be 
understood qualitatively using a combination of the mean-motion
Hamiltonian, equation~\ref{eq:swit01} below, and the resonance Hamiltonian. We
assume that initially the system is in a given $\Ie$-state with its
conjugate variable uniformly distributed in $(0,\pi)$. This initial phase
line, $\sC_{0}$, evolves during the switch-on period in the mean-motion
Hamiltonian: in scaled units, with $\In=1$, this is
\drlabel{eq:swit01}
\begin{equation}
\Kbb = \frac{3\lambda(t)^{2}}{16}
\left( \Fs^{2}+\frac12\Fm^{2}\right)\Ie^{2} +
\frac14\lambda(t)\omo\Fm A(\Ie)B(\Ie)\sin\left( 2\psie-3g(t)\right)\sin\omo t,
\label{eq:swit01} 
\end{equation}
where
\(
g(t) = \int_{0}^{t}dt\, \lambda(t) \left( \Fs+\Fm\cos\omo t\right)
\).
For $t\geq T_{a}$ (and before the switch-off time), $\lambda=1$ and 
this Hamiltonian simplifies to
\[
\Kbb=\frac{3}{16}\left( \Fs^{2}+\frac12\Fm^{2}\right)\Ie^{2} +
\frac14\omo\Fm A(\Ie)B(\Ie)\sin\left( 2\psie-3g_{a}-3\Fs t+
\frac{3\Fm}{\omo}\sin\omo t\right)\sin\omo t
\]
where
\(
g_{a}=g(T_{a})-T_{a}\Fs - \frac{\Fm}{\omo}\sin\omo T_{a}
\).
Near the $j$th resonance this can be approximated by the resonance
Hamiltonian,
\drlabel{eq:swit02}
\begin{equation}
K_{j}=\frac{3}{16}\left( \Fs^{2}+\frac12\Fm^{2}\right)
\left( \Ie-\alpha_{j}\right)^{2} +\frac12\omo\Fm A(\Ie)B(\Ie)
J'_{j}\left( \frac{3\Fm}{\omo}\right) \cos(2\thetae + j\pi),
\label{eq:swit02}
\end{equation}
where $\alpha_{j}$ is defined after equation~\ref{eq:st20a} and 
$2\thetae=2\psie-(3\Fs-j\omo)t-3g_{a}$, ($t\geq T_{a}$).

During the period $0\leq t \leq T_{a}$ the initial phase curve evolves 
according to the Hamiltonian~\ref{eq:swit01} into the line $\sC_{a}$. Hence
by plotting the line $\sC_{a}$ and the contours of $K_{j}$ 
(the dashed lines), we obtain a 
qualitative picture showing how $\Pion$ can be affected by the field switch.

In the following three figures we show the separatrix of $K_{j}$ and the line
$\sC_{a}$ for $\Fs=0.0263$, $\Fm=0.14$, $\omo=0.0114$, $\Ism=0.2$ 
and the initial state
$\Ie(0)=-0.3$, when $N_{a}=1$, 3 and 35: the field values are slightly 
different from those used to generate figures~\ref{f:18a} and~\ref{f:18b} 
because an approximate Hamiltonian is used.

\vspace{12pt}

\begin{center}
\drlabel{f:separ}
\begin{drfigs}
\drsetfig{320pt}{separatrix.eps}{
Diagram showing the curves $\sC_{a}$, formed by evolving the inital
phase curve $\Ie(0)=-0.3$ through various switch times, 
$T_{a}=2\pi N_{a}/\Omega$. The dashed lines are the separatrixes of the 
resonance hamiltonian, equation~\ref{eq:swit02}, at time $T_{a}$.}{f:separ}
\end{drfigs}
\end{center}

For $N_{a}=1$ the line $\sC_{a}$ is close to the initial line, $\Ie(0)=-0.3$.
Slightly less than half of $\sC_{a}$ lies inside the separatrix and these
orbits will be transported to regions of larger $\Ie$ and some will ionise
depending upon the value of $\Ie^{c}$.

For $N_{a}=3$ the curve $\sC_{a}$ is more distorted; a smaller proportion 
of orbits lie inside the separatrix, but all of these are close to it and
will all be transported to larger $\Ie$ than the equivalent points of the 
previous example. 

For $N_{a}=35$, $\sC_{a}$ has developed a complicated shape due to the 
motion inside the island. In this example a significant proportion of the 
orbits inside the separatrix are close to the horizontal line through the 
island centre, so will not be transported to regions of large $\Ie$. Clearly
this structure is very sensitive to changes in $N_{a}$ and this sensitivity
will be reflected in $\Pion$.

These figures provide a qualitative explanation for the complications seen
in figures~\ref{f:18a} and~\ref{f:18b}. In particular they show why there is
no simple, precise relation between $\Fs^{(j)}$ and $\sF_{s}^{(j)}$; 
they also show that the dynamics underlying
the apparently simple resonances seen in figure~\ref{f:1} is very complicated.

\section{Conclusions}
In this paper we examine the behaviour of a classical hydrogen atom in 
parallel static and microwave fields, with  frequencies that are low
by comparison to the unperturbed orbital frequency. There are three main
reason why the  classical atom is considered. 

First, it is not 
necessary to make dynamical approximations in order to numerically 
integrate the classical equations of motion. The errors of the estimated
ionisation probability are determined mainly by the Monte-Carlo
sampling errors and with modern computers these can be made acceptably small.
This is in contrast to quantal calculations in which unquantifiable 
approximations have to made in order to solve the corresponding equations
of motion.

Second, within the framework of classical dynamics there is a range of 
easily applied approximations that help provide understanding of observed
phenomena. The corresponding approximations are not so easy to apply
to either Schr\"{o}dinger's of Heisenberg's equations of motion.

Finally, using techniques of analytical dynamics it is possible
to construct an approximate Hamiltonian, which provides a fairly accurate
approximation to the exact classical dynamics, see figure~\ref{f:8}, and 
which may be used as a basis for feasible quantal calculations.

The main effects of interest here are the resonances between the microwave
field and the Stark frequency induced by the static field. These resonances
were first observed by Galvez {\em et al} (2000) and this paper also presents 
the first theory to describe these resonances qualitatively. It was shown 
how these resonances are responsible for an enhanced ionisation signal over 
a narrow range of static field 
strengths, for fixed microwave field amplitude and frequency. Additionally
these signals disappear at particular combinations of the two field 
amplitudes and recent experiments, Schultz (2003), Galvez {\em et al} (2004) 
and Schultz {\em et al} (2004), have extended the measurements of these
``disappearance fields''.

Since the first observations of these resonances three theoretical papers
describing the phenomena from different perspectives have been published.
Oks and Uzer (2000) use a Floquet analysis to derive zero-order estimates
of $\Fs^{(j)}$ and $\Fm^{(j,k)}$. Robicheaux {\em et al} (2002) have solved
Schr\"{o}dinger's equation  for this problem using a split-operator method
and have made comparisons with the classical and experimental 
ionisation probabilities for $\nze=39$ across the $j=1$ resonance with 
$\Fm=0.144$. These calculations suggest that the classical and quantal 
values of $\sF_{s}^{(1)}$ are very close, but that quantal value of $\Pion$ 
is smaller than the classical value for $\Fs>\sF_{s}^{(1)}$. In addition the
time-dependence of the ionisation probability is described for three values
of $\Fs$ and this appears to contradict the results summarised in 
figure~\ref{f:time}, though it is difficult to make comparisons between 
substate averaged and unaveraged data. Ostrovsky and Horsdal-Pedersen (2003) 
use an energy shell subspace with a time-dependent electric field 
and weak, perpendicular
magnetic field, with the aim of understanding oscillations seen in the
experimental result, subsequently attributed to another cause, see 
Wilson {\em et al} (2004). This analysis is based on the same type of 
averaging 
approximation that leads to the Hamiltonian~\ref{eq:st15}, see also 
Born (1960, section~38), and inevitably gives zero-order estimates
of $\Fs^{(j)}$ and $\Fm^{(j,k)}$.

In this paper we have described the classical dynamics of this system in 
more detail: in particular more accurate values of $\Fs^{(j)}$ and 
$\Fm^{(j,k)}$ are determined and the properties of the classical 
resonances are described in some detail. We now list these features 
and discuss the probable consequences to the quantum mechanics.

\newcounter{conc}
\begin{list}%
{\arabic{conc})}{\usecounter{conc}%
\setlength{\labelsep}{5pt}\setlength{\leftmargin}{10pt}%
\setlength{\labelwidth}{0pt}%
\setlength{\listparindent}{0pt}}

\item \label{note:conc1}We have established that the position 
of the dynamical resonance,
 $\Fs^{(j)}$, as computed by theory, is not at precisely the field
$\sF_{s}^{(j)}$, at which the ionisation probability is largest, although
$\Fs^{(j)}\simeq \sF_{s}^{(j)}$. Further, we have shown, numerically, that
the difference $\Fs^{(j)}- \sF_{s}^{(j)}$ can depend on the field
envelope.

\item \label{note:conc2}We have isolated the terms in the Hamiltonian 
that give rise to
the dynamical resonance. This allows the computation of the 
dynamical resonance position using high-order perturbation theory; where this 
series converges we obtain improved estimates of the resonance position.

This analysis is essentially the same as the corresponding quantal
theory which we therefore expect to give the same result. Further, this
suggests that the discrepancy noted in~(\ref{note:conc1}) above will
also occur in an accurate quantal calculation.

\item \label{note:conc3}Using a classical approximation, based on 
two stages of averaging,
we have derived a number of conditions necessary for the dynamical 
resonance to affect the ionisation probability. These depend upon properties
of the classical resonance island, the most significant being the island 
width. 

A dynamical resonance affects the ionisation probability only if it is wide
enough to bridge the gap in phase space between the initial state 
and those states that ionise, see section~\ref{sec:res-ion-II}. Because 
the ratio of these two actions is independent of the initial principal
number, we expect a similar story in the 
quantal description, though quantal effects will inevitably blur these
boundaries. This suggests that there may be cases where a resonance not
seen in the classical ionisation probabilities will be visible in the 
quantal probabilities.
				   
Besides the island width, its area also plays a role in quantum mechanics; 
this area is proportional to the initial principal quantum 
number, $\nze$, so we expect the resonances seen here and in current 
experiments to change, and possibly disappear, as $\nze$ decreases.

\item \label{note:conc4}The resonances in the ionisation probability, 
not averaged over substates, are generally
very sharp, see for instance figure~\ref{f:n21a} and table~\ref{t:10}. The
full-width of an isolated resonance is generally smaller than its 
theoretical shift produced by changing $\Ism=m\hbar$, as seen by comparing the
data presented in tables~\ref{t:10} and~\ref{t:11}.  However, resonances
near an adiabatic boundary are significantly broadened and asymmetries
are introduced, figure~\ref{f:w03} and~\ref{f:w04}. This causes 
substate-averaged resonances to be far wider than isolated resonances, and 
also affects their shape. 

The classical dynamics of this process is complicated and not understood. 
Because the experimentally observed resonances behave in a similar fashion, we 
expect a similar dynamical effect in any quantal calculation, but understanding
this is harder than understanding the classical dynamics.

\item \label{note:conc5}The shape of an isolated classical resonance 
and the value of
$\sF_{s}^{(j)}$, with no substate averaging, can be affected by the field 
envelope, if the fields are switched on sufficiently slowly, see 
figures~\ref{f:18a} and~\ref{f:18b}. These changes are caused by the phase 
line representing the initial state becoming tangled as the separatrix of
the resonance island passes through it, see figure~\ref{f:separ}. We expect 
this behaviour to be seen in the quantum dynamics provided the principal
quantum number is large enough, but how large is not known.

\item \label{note:conc6}Classical resonances develop over a time scale that 
is much longer than that 
of the ionisation process operating away from resonance. This is 
because, at resonance, ionisation happens by transport around the
resonance island and this is a relatively slow process. Further, across
a resonances the ionisation time appears to reflect the island dynamics. For
instance, at the edge of a resonance the ionisation time is longest because
motion near the separatrix is very slow: this and other, not 
understood, features are seen in the lower panel of figure~\ref{f:time}.

As with point~\ref{note:conc5} above, we should expect to see similar behaviour
in a quantal calculation, provided the principal quantum number is sufficiently
large, but how large is not known.

\end{list}

\subsubsection*{Acknowledgements}
It is a pleasure to thank Professor Peter Koch for valuable discussions 
and encouragement during all stages of this work,
and for helpful comments on drafts of this manuscript.

\section{Appendix: action variables}
The derivation of the required results is easiest if scaled variables
are used. If $a$ is any scale length and $\Ize$ an action, suitable scaled 
variables are
\[
\tilde{F}=\frac{\mu a^{3}}{\Ize^{2}}F, \quad 
\tilde{E}=\frac{\mu a^{2}}{\Ize^{2}}E, \quad
\tilde{\alpha}=\frac{\mu e^{2}a}{\Ize^{2}}\alpha, \quad
\tilde{I}_{k}=\frac{I_{k}}{\Ize}.
\]
Taking $\Ize$ to be the initial value of $\In$ and $a$ the semi-major axis 
of the initial Kepler ellipse, $a=\In^{2}/\mu e^{2}$ we have
\[
\tilde{F}=\frac{\Ize^{4}}{\mu^{2}e^{6}}F, \quad
\tilde{E}=\frac{\Ize}{\mu e^{4}}E, \quad
\tilde{\alpha}=\alpha.
\]
In the following these scaled units are used but for clarity the tilde
is not shown. Most of these results are obtained using a Maple program
to manipulate the series, which were computed to higher-order than
quoted here.

\subsubsection*{Series for $\Ion$}
For $\Ion$ set $\xi^{2}=y$ so equation~\ref{eq:2-08a} becomes
\drlabel{eq:st-a01}
\begin{equation}
\Ion=\frac{1}{2\pi} \int_{y_{1}}^{y_{2}} \frac{dy}{y} \sqrt{f_{1}(y)}, \quad
f_{1}(y)=-Fy^{3}-2Wy^{2}+2\alpha_{1}y -\Ism^{2},
\label{eq:st-a01}
\end{equation}
where $W=-E>0$.
In the parameter range of interest $f_{1}(y)$ has three real roots, two 
positive $0\leq y_{1}<y_{2}$ (with $y_{1}=0$ only when $\Ism=0$) and one 
negative root, $y_{3}=-y_{-}/F < 0$. We may write
\[
f_{1}(y)=(Fy+y_{-})(y-y_{1})(y_{2}-y), 
\quad y_{1}y_{2}y_{-}=\Ism^{2}, \quad (y_{1}+y_{2})F=y_{-}-2W,
\]
so that $y_{-}=O(1)$ and, if $F=0$, $y_{-}=2W$. Only the 
value of $y_{-}$ is needed because all quantities of interest can be 
expressed in terms of the combinations $y_{1}+y_{2}$ and $y_{1}y_{2}$. 
Then the series for $\Ion$ is
\drlabel{eq:st-a01a}
\begin{equation}
\Ion = \sqrt{y_{-}}  \int_{y_{1}}^{y_{2}} \frac{dy}{y}
 \sqrt{(y-y_{1})(y_{2}-y)} \sqrt{1+\frac{Fy}{y_{-}}},
= \sqrt{y_{-}} \sum_{k=0}^{\infty} a_{k} 
\left( \frac{F}{y_{-}} \right)^{k} H_{k},\label{eq:st-a01a}
\end{equation}
where 
\[
H_{k}=\frac{1}{4\pi} \oint_{\sC} dz\, z^{k-1}  h(z)
\quad {\rm and} \quad
a_{k}=\frac{\sqrt{\pi}}{2\, k! \,\Gamma(3/2-k)},
\]
and where $h(z)=-\sqrt{(z-y_{1})(y_{2}-z)}$ has a cut on the real
axis between $y_{1}$ and $y_{2}$ such that on the real axis between these two
roots on the upper branch $h(x)<0$, and on
the lower branch $h >0$; for $x$ real and $x>y_{2}$,
$h(x)=i\sqrt{(x-y_{1})(x-y_{2})}$ and for $x<y_{1}$,
$h(x)=-i\sqrt{(y_{1}-x)(y_{2}-x)}$. The 
contour $\sC$  encloses the branch cut between $y_{1}$ and 
$y_{2}$ but not the origin. 

If $k=0$ the integrand has a pole at $z=0$ and contributions from the
circle $z=Re^{i\theta}$, as $R\to\infty$, so
\[
H_{0}=\frac14 (\sqrt{y_{2}}-\sqrt{y_{1}})^{2}.
\]
For $k\geq 1$  the only contribution is from the circle at infinity, 
$\sC_{\infty}$ 
\[
H_{k} =\frac{i}{4\pi} \oint_{\sC_{\infty}} dz\, z^{k} 
\sqrt{1-\frac{y_{1}}{z}}\sqrt{1-\frac{y_{2}}{z}}
= \frac{(-1)^{k}}{2} \sum_{r=0}^{k+1} 
a_{r}a_{k+1-r} \,y_{1}^{r}\,y_{2}^{k+1-r},  \quad k\geq 1.
\]
Some values are 
\begin{eqnarray*}
H_{1} &=& \frac{1}{16} (y_{1}+y_{2})^{2} -\frac14 y_{1}y_{2},  \quad
H_{2}= \frac{1}{32}  (y_{1}+y_{2})\left[  (y_{1}+y_{2})^{2} - 
4y_{1}y_{2}\right],\\
H_{3} &=& \frac{1}{256} \left[ (y_{1}+y_{2})^{2} - 4y_{1}y_{2} \right] 
\left[ 5(y_{1}+y_{2})^{2} - 4y_{1}y_{2}\right], \\
H_{4} &=& \frac{1}{512} \left( y_{1}+y_{2}\right)
\left[ (y_{1}+y_{2})^{2}-4y_{1}y_{2}\right]
\left[ 7(y_{1}+y_{2})^{2}-12y_{1}y_{2}\right].
\end{eqnarray*}

\subsubsection*{Perturbation expansion for $y_{-}$}
Now we need an expression for $y_{-}>0$ which is proportional to the 
negative root of $f_{1}(-y_{-}/F)=0$. Because $y_{-}=O(1)$, it is 
expedient to define $y=-zF$ to give the equation
\[
z^{3} - 2Wz^{2}-2\alpha_{1} F z - F^{2}\Ism^{2}=0
\]
When $F=0$, $z=2W$, so we put $z=2W + z_{1}F + \cdots$,
then perturbation theory gives
\[
z=y_{-}=2W + \frac{\alpha_{1}F}{W} + \frac{F^{2}}{4W^{3}} 
\left(W\Ism^{2}-2\alpha_{1}^{2} \right)  - \frac{\alpha_{1}F^{3}}{8W^{5}} 
\left( 3W\Ism^{2}-4\alpha_{1}^{2} \right) +\cdots\,.
\]
Thus 
\[
y_{1}+y_{2}
= \frac{\alpha_{1}}{W} + \frac{F}{4W^{3}} \left(W\Ism^{2}-2\alpha_{1}^{2} 
\right)  - \frac{\alpha_{1}F^{2}}{8W^{5}} \left( 3W\Ism^{2}-4\alpha_{1}^{2} 
\right) +\cdots.
\]
and 
\[
y_{1}y_{2} = \frac{\Ism^{2}}{2W} - \frac{\alpha_{1}\Ism^{2}}{4W^{3}}F -
\frac{\Ism^{2}}{16W^{5}} \left( W\Ism^{2}-4\alpha_{1}^{2} \right) F^{2} + 
\cdots\,.
\]
Substituting these expressions for $y_{1}+y_{2}$ and $y_{1}y_{2}$ into
the above expressions for $H_{k}$ gives
\drlabel{eq:a02}
\begin{equation}
\Ion = \frac{\alpha_{1}}{2\sqrt{2W}} +\frac12\Ism - 
\frac{\sqrt{2}F}{64W^{5/2}} \left( 2\Ism^{2}W -3\alpha_{1}^{2}\right) 
- \frac{5\sqrt{2}\alpha_{1}F^{2}}{1024W^{9/2}} 
\left( 6\Ism^{2}W-7\alpha_{1}^{2}\right)+ \cdots.
\label{eq:a02}
\end{equation}

\subsubsection*{Series for $\Itw$}
For $\Itw$ we set $\eta^{2}=u$, and equation~\ref{eq:2-08b} gives
\[
\Itw=\frac{1}{2\pi}\int_{u_{1}}^{u_{2}}\frac{du}{u}\,\sqrt{f_{2}(u)},
\quad f_{2}(u)=Fu^{3}-2Wu^{2}+2\alpha_{2}u-\Ism^{2}.
\]
In the parameter range of interest $f_{2}$ has three real positive roots, 
$0\leq u_{1} \leq u_{2}$ and $u_{+}/F$, so we write
\(
f_{2}(u)=(u_{+}-Fu)(u-u_{1})(u_{2}-u)
\)
with $u_{1}u_{2}u_{+}=\Ism^{2}$ and $u_{1}+u_{2}=(2W-u_{+})/F$. Thus
\begin{equation}
\Itw = \frac{\sqrt{u_{+}}}{2\pi} \int_{u_{1}}^{u_{2}} \frac{du}{u} 
\sqrt{(u-u_{1})(u_{2}-u)}\sqrt{1-\frac{uF}{u_{+}}}
= \sqrt{u_{+}}\sum_{k=0}^{\infty} a_{k} \left(-\frac{F}{u_{+}} \right)^{k} 
H_{k}
\end{equation}
where $a_{k}$ and $H_{k}$ are defined above.

The perturbation expansion for $u_{+}$ is given, as before, by setting 
$u=z/F$ to write the equation $f_{2}=0$ in the form
\(
z^{3} - 2Wz^{2}+ 2\alpha_{2} F z - F^{2}\Ism^{2}=0,
\)
giving
\[
u_{+}=2W - \frac{\alpha_{2}F}{W} + \frac{F^{2}}{4W^{3}} 
\left(W\Ism^{2}-2\alpha_{2}^{2} \right)  + 
\frac{\alpha_{2}F^{3}}{8W^{5}} \left( 3W\Ism^{2}-4\alpha_{2}^{2} \right) +
\cdots\,.
\]
This then gives
\[
u_{1}+u_{2}
= \frac{\alpha_{2}}{W} - \frac{F}{4W^{3}} \left(W\Ism^{2}-2\alpha_{2}^{2} 
\right)  - \frac{\alpha_{2}F^{2}}{8W^{5}} \left( 3W\Ism^{2}-4\alpha_{2}^{2} 
\right) +\cdots.
\]
and 
\[
u_{1}u_{2} 
= \frac{\Ism^{2}}{2W} + \frac{\alpha_{2}\Ism^{2}}{4W^{3}}F -
\frac{\Ism^{2}}{16W^{5}} \left( W\Ism^{2} - 4\alpha_{1}^{2}\right) F^{2} + 
\cdots
\]
and hence the expression for $\Itw$ is,
\[
\Itw=\frac{\alpha_{2}}{2\sqrt{2W}} -\frac12 \Ism 
-\frac{\sqrt{2}F}{64W^{5/2}}\left( 2\Ism^{2}W-3\alpha_{2}^{2}\right)
-\frac{5\sqrt{2}\alpha_{2}F^{2}}{1024W^{9/2}}
\left( 6\Ism^{2}W-7\alpha_{2}^{2}\right)+\cdots.
\]

These series for $\Ion$ and $\Itw$ now need to be inverted to give
$\alpha_{1}$, $\alpha_{2}=2-\alpha_{1}$ and $W$ as power series in $F$. The
zero-order term is trivial so we substitute the series
\[
\alpha_{1} = \sigma_{1} + \sum_{k=1}^{\infty}c_{k}F^{k},
\quad \sigma_{1}=\frac{2\Ion+\Ism}{\In},  \qquad
W = \frac{1}{2\In^{2}}+\sum_{k=1}^{\infty}W_{k}F^{k},
\]
into the series for $\Ion$ and $\Itw$ and solve for the unknown coefficients to
give the energy and the separation constant
quoted in equations~\ref{eq:2-09} and~\ref{eq:2-09a}. 

\subsection{Angle variables}
Using the definition $\theta_{k}=\cd S/\cd I_{k}$ and the notation introduced
in the previous section,
\drlabel{eq:app-ang01}
\begin{equation}
\theta_{k}=\frac14 \int_{y_{1}}^{y}\frac{dy}{y} 
\frac{\cd f_{1}/\cd I_{k}}{\sqrt{f_{1}(y)}} +
\frac14 \int_{u_{1}}^{u}\frac{du}{u} 
\frac{\cd f_{2}/\cd I_{k}}{\sqrt{f_{2}(u)}}.
\label{eq:app-ang01}
\end{equation}
By differentiating equations~\ref{eq:2-08a} and~\ref{eq:2-08b} we see that
\drlabel{eq:app-ang02}
\begin{equation}
\int_{y_{1}}^{y_{2}}\frac{dy}{y} 
\frac{\cd f_{1}/\cd I_{k}}{\sqrt{f_{1}(y)}}=2\pi\delta_{1k}
\quad {\rm and} \quad 
\int_{u_{1}}^{u_{2}}\frac{du}{u} 
\frac{\cd f_{2}/\cd I_{k}}{\sqrt{f_{2}(u)}}=2\pi\delta_{2k}.
\label{eq:app-ang02}
\end{equation}
Using the factorisation introduced in the previous section and putting
\drlabel{eq:app-ang02a}
\begin{equation}
y=y_{1}\cos^{2}(\psi/2)+y_{2}\sin^{2}(\psi/2), \quad
u=u_{1}\cos^{2}(\chi/2)+u_{2}\sin^{2}(\chi/2),
\label{eq:app-ang02a}
\end{equation}
we obtain
\drlabel{eq:app-ang03}
\begin{eqnarray}
\theta_{k} &=& \frac{1}{2\sqrt{y_{-}}}\int_{0}^{\psi}d\psi\,
\left( \vsp{10}\alpha_{1k}-W_{k}y(\psi)\right)
\left(1+\frac{Fy(\psi)}{y_{-}} \right)^{-1/2},\nonumber \\
&& \hspace{20pt} -\frac{1}{2\sqrt{u_{+}}}\int_{0}^{\chi}d\chi\,
\left( \vsp{10}\alpha_{1k}+W_{k}u(\chi)\right)
\left(1-\frac{Fu(\chi)}{u_{+}} \right)^{-1/2},
\label{eq:app-ang03}
\end{eqnarray}
where $\alpha_{1k}=\cd \alpha_{1}/\cd I_{k}$ and $W_{k}=\cd W/\cd I_{k}$; the
series expansions for both these variables are derived in the previous section.
The first integrand can be expressed as series in $\cos n\psi$ and the 
second as a series in $\cos n\chi$, so integration gives
\drlabel{eq:app-ang04}
\begin{equation} 
\theta_{k} = c_{k1}\psi + P_{k}(\psi) + c_{k2}\chi + Q_{k}(\chi), \quad
k=1,\,2,
\label{eq:app-ang04}
\end{equation}
where $(P_{k}(x),Q_{k}(x))$ are odd $2\pi$-periodic functions. Using the 
relations~\ref{eq:app-ang02} we see that $c_{11}=c_{22}=1$ and 
$c_{12}=c_{21}=0$ and evaluation of these integrals, to $O(F)$, gives their
Fourier series representations
\drlabel{eq:app-ang05}
\begin{eqnarray}
4P_{1} &=& -\sigma_{1}\left[ \vsp{10} 4-
\In^{3}(2\In+7(2\Itw+\Ism))F\right]\sin\psi
-\In^{4}\sigma_{1}^{2} F\sin 2\psi, \nonumber \\
4P_{2} &=& -\sigma_{1}\left[ \vsp{11} 4-
\In^{3}(10\In+7(2\Itw+\Ism))F\right]\sin\psi
-\In^{4}\sigma_{1}^{2}F\sin 2\psi, \label{eq:app-ang05} \\
4Q_{1} &=& -\sigma_{2}\left[ \vsp{11} 4 +
\In^{3}(10\In+7(2\Ion+\Ism))F\right]\sin\chi +
 \In^{4}\sigma_{2}^{2}F\sin 2\chi, \nonumber \\
4Q_{2} &=& -\sigma_{2}\left[ \vsp{11} 4 +
\In^{3}(2\In+7(2\Ion+\Ism))F\right]\sin\chi
+\In^{4}\sigma_{2}^{2}F\sin 2\chi.  \nonumber
\end{eqnarray}
where $\sigma_{k}=\sqrt{I_{k}(I_{k}+\Ism)}/\In$, $k=1,\,2$. When $F=0$ these
expression lead  to the formulae quoted in equation~\ref{eq:2-10}.

\subsection{Evaluation of $\cd S/\cd F$}
The generating function $S_{1}(\thetaon,\thetatw,\xi,\eta)$ returns
to its initial value when either $\xi$ or $\eta$ increases through
a period, see Born (1960, page~82); here we use the notation 
of Goldstein (1980) to label generating functions. It follows that the 
Hamiltonian~\ref{eq:2-11} is a periodic function of the angle variables with
zero mean value. By differentiating the generating function $S(\bI,\xi,\eta)$
with respect to $F$ and using the angles $(\psi,\chi)$, defined in 
equations~\ref{eq:app-ang02a} we see that $\cd S/\cd F$ may be written
in the form
\drlabel{eq:app-act01}
\begin{eqnarray}
\pad{S}{F} &=& \frac{1}{4\sqrt{y_{-}}}\int_{0}^{\psi}d\psi\, 
\left( -y^{2}-2W_{F}y+2\alpha_{1F}\right) 
\left( 1+\frac{Fy}{y_{-}}\right)^{-1/2} \nonumber \\
&& \hspace{20pt}+\frac{1}{4\sqrt{u_{+}}}\int_{0}^{\chi}d\chi\, 
\left( u^{2}-2W_{F}u-2\alpha_{1F}\right) 
\left( 1-\frac{Fu}{u_{+}}\right)^{-1/2}, \label{eq:app-act01}
\end{eqnarray}
where $W_{F}=\cd W/\cd F$ and $\alpha_{1F}=\cd \alpha_{1}/\cd F$. The two
integrands can be expressed in terms of even Fourier series in
$\psi$ and $\chi$ respectively
where the constant terms are missing because of the argument given at the
beginning of this section and because of the general result 
$\cd S/\cd t=\cd S_{1}/\cd t$, which gives $\cd S/\cd F=\cd S_{1}/\cd F$.
Hence the integral~\ref{eq:app-act01} leads to the Fourier series
defined in equation~\ref{eq:2-10b},
where the coefficients $(A_{k},B_{k})$ depend upon $\Fs$, $\Fm$ and the
action variables. Because of relation~\ref{eq:app-ang04} it follows that no
term of $\dot{F}\cd S/\cd F$ is independent of the angle variables.

The Fourier series representation of $\cd S/\cd F$ is obtained in the same
manner as that for $P_{k}(\psi)$ and $Q_{k}(\chi)$: to $O(F)$, we have
\begin{eqnarray*}
A_{1} &=& \frac12\In^{4}\sigma_{1} \left\{ \vsp{12}
(\Ion+3\Itw+2\Ism) \right. \\
&& \hspace{30pt} \left. \vsp{12}
-\frac14\In^{3}F\left( 14\Ion^{2}+27\Ion\Ism+26\Ion\Itw+22\Itw^{2}+
35\Itw\Ism+22\Ism^{2}\right) \right\},  \\
A_{2} &=& -\frac14\In^{5}\sigma_{1}^{2} 
\left\{ \vsp{10} 1 - F\In^{3} (4\Ion+6\Itw+5\Ism )\right\},\quad
A_{3} = -\frac{1}{12} F\In^{9}\sigma_{1}^{3}, \\
B_{1} &=& -\frac12\In^{4}\sigma_{2} \left\{ \vsp{12}
\left( 3\Ion + \Itw+2\Ism\right)  \right. \\
&& \hspace{30pt} \left. \vsp{12}
-\frac14\In^{3}F \left( 22\Ion^{2}+35\Ion\Ism+26\Ion\Itw +2\Itw^{2} +
15\Itw\Ism + 22\Ism^{2}\right) \right\}, \\
B_{2} &=& \frac14\In^{5}\sigma_{2}^{2} \left\{ 1+
F\In^{3}\left(3\Ion+7\Itw +5\Ism \right)\right\},\quad
B_{3} = -\frac{1}{12}F\In^{9}\sigma_{2}^{3}.
\end{eqnarray*}

\newpage
\section*{References}
\begin{trivlist}
\item[] Abrines R and Percival I C 1966 Proc Phys Soc {\bf 88} 861-72 
\item[] Banks D and Leopold J G  Stark 1978 J Phys {\bf B11} 2833-43
\item[] Bayfield J and Koch P M 1974 Phys. Rev Lett {\bf 33} 258-??
\item[] Bellermann M R W, Koch P M, Mariani D R and Richards D 1996 Phys 
Rev Lett {\bf 76} 892-5
\item[] Bellermann M R W, Koch P M and Richards D 1997 Phys Rev Lett {\bf 78}
3840-3
\item[] Bethe H A and Salpeter E E  1957 {\em Quantum mechanics of One- and
two-Electron Systems}, Handbuch der Physik Ed S Fl\"{u}gge (Springer-Verlag)
\item[] Born M 1960 {\em The Mechanics of the Atom} (Frederick Ungar 
Publishing Co, New York)
\item[] Casati J, Chirikov B V and Shepelyansky D L 1984 Phys Rev Lett
{\bf 53} 2525-8
\item[] Damburg R J and Kolosov V V 1983 {\em Hydrogen Rydberg atoms in
electric fields} in Rydberg States of atoms and molecules, Eds  
R F Stebbings and F B Dunning (Cambridge University Press)
\item[] Dando P A and Richards D 1993 J Phys {\bf B26} 3001-16
\item[] Delone N B, Zon B A and Krainov V P 1978 Sov. Phys. JETP {\bf 48} 223 
[Zh. Eksp. Teor. Fiz {\bf 75} 445 (1978)]
\item[] Goldstein H 1980 {\em Classical Dynamics} (Addison-Wesley)
\item[] Howard J E 1995 Phys Rev {\bf A51} 3934-46
\item[] Jensen R V, Susskind S M and Sanders MM 1991 {\em Chaotic Ionization
of Highly Excited Hydrogen Hydrogen Atoms}, Phys Rep {\bf 201} 1-56
\item[] Koch P M 1990 {\em Microwave ionization of excited hydrogen atoms:
What we do and do not understand}, in Chaos, pages 441-75, Ed D K Campbell 
(Amer. Inst. Phys, New York)
\item[] Koch P M 1995 Physica {\bf D83} 178-205
\item[] Koch P M and Bellermann M R W 2000 Physica {\bf A288} 98-118
\item[] Koch P M and van Leeuwen 1995 Phys Rep {\bf 255} (5 and 6) 289-405
\item[] Leopold J and Percival I C 1979 Phys Rev Lett {\bf 41} 944-7
\item[] Leopold J and Richards D 1991 J Phys {\bf B24} 1209-40
\item[] Leopold J and Richards D 1994 J Phys {\bf B27} 2169-89
\item[] Galvez E L, Sauer B E, Moorman L, Koch P M and Richards D 1988,
Phys Rev Lett {\bf 61} 2011-4
\item[] Galvez E K, Koch P M, Richards D and Zelazny S A 2000 Phys Rev 
{\bf A61} 060101(4)
\item[] Galvez E K, Wislon J W, Schultz K D, Koch P M and  Richards D 
2004 {\em Resonances in high $n$ H atoms ionized by collinear static and 
microwave fields}, in preparation.
\item[] Hammersley J M and Handscomb D C (1964) {\em Monte Carlo Methods},
Methuen (Monograph on Applied Probability and Statistics)
\item[] Oks E and Uzer T  2000  J Phys {\bf B23} 1985-95
\item[] Ostrovsky V N  and Horsdal-Pedersen E 2003 Eur Phys J {\bf D23}
15-24
\item[] Meerson B I, Oks E A and Sasorov P V 1979 Sov Phys JETP Lett {\bf 29}
72
\item[] Rath O and Richards D 1988 J Phys {\bf B21} 555-71
\item[] Rath O 1990 {\em The dynamics of excited hydrogen atoms in 
strong electric and Magnetic fields}, PhD Thesis (Open University)
\item[] Richards D 1987  J Phys {\bf B20} 2171-92
\item[] Richards D, Leopold J G, Koch P M, Galvez E J, can Leeuwen K A H,
Moorman K, Sauer B E and Jensen R V 1989 J Phys {\bf B22} 1307-33
\item[] Richards D 1996a J Phys {\bf B29} 2925-49
\item[] Richards D 1996b J Phys {\bf B29} 5237-71
\item[] Richards D 1997 J Phys {\bf B30} 4019-47
\item[] Robicheaux F, Oks E, Parker A L and Uzer T 2002 J Phys {\bf B35} 4613-8
\item[] Sauer B E, Yoakum S, Moorman L, Koch P M, Richards D and Dando P A 
1992 Phys Rev Lett {\bf 68} 468-71
\item[] Schultz K D 2003 {\em Resonances in H atoms in Collinear Linearly
Polarized Microwave Field and Static Field}, PhD Thesis, State University
of New York at Stony Brook, Dept. of Physics.
\item[] Schultz K D, Koch P M and Richards D  2004 in preparation
\item[] Silverstone H J 1978 Phys Rev lett {\bf A18} 1853-64
\item[] Szebehely V 1967 {\em Theory of orbits} (Academic Press)
\item[] van Leeuwen K A H, v. Oppen G, Renwick S, Bowlin J B, Koch P M,
Jensen R V, Rath O, Richards D and Leopold J G 1985 Phys Rev Lett {\bf 55}
2231-4.
\item[] Wilson J W 2003 {\em Polarization Dependence of Microwave Ionization
at High Scaled Frequencies}, PhD Thesis, State University of New York at 
Stony Brook, Dept. of Physics.
\item Wilson J W, Schultz K D, Walter D and Koch P M 2004 {\em Experimental 
control of oscillations in ionization of excited hydrogen atoms
by collinear microwave and static fields}, submitted to J Phys {\bf B}
\end{trivlist}

\end{document}